%% file: main__1_.tex
\documentclass[sigconf]{aamas}

\usepackage{amsmath,amssymb,amsfonts}

\usepackage{algorithm}
\usepackage{algpseudocode}
\usepackage{graphicx}    
\usepackage{subcaption}  

\usepackage{balance} 
\newtheorem{lemma}{Lemma}
\newtheorem{assumption}{Assumption}

\newtheorem{theorem}{Theorem}

\usepackage{balance} 

\doi{ARUH6291}

\makeatletter
\gdef\@copyrightpermission{
  \begin{minipage}{0.2\columnwidth}
   \href{https://creativecommons.org/licenses/by/4.0/}{\includegraphics[width=0.90\textwidth]{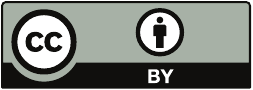}}
  \end{minipage}\hfill
  \begin{minipage}{0.8\columnwidth}
   \href{https://creativecommons.org/licenses/by/4.0/}{This work is licensed under a Creative Commons Attribution International 4.0 License.}
  \end{minipage}
  \vspace{5pt}
}
\makeatother

\setcopyright{ifaamas}
\acmConference[AAMAS '26]{Proc.\@ of the 25th International Conference
on Autonomous Agents and Multiagent Systems (AAMAS 2026)}{May 25 -- 29, 2026}
{Paphos, Cyprus}{C.~Amato, L.~Dennis, V.~Mascardi, J.~Thangarajah (eds.)}
\copyrightyear{2026}
\acmYear{2026}
\acmDOI{}
\acmPrice{}
\acmISBN{}

\title[Regularization-free Last-iterate Convergence in Zero-sum Games via BNN Dynamics]{Teaching an Old Dynamics New Tricks: Regularization-free Last-iterate Convergence in Zero-sum Games via BNN Dynamics}

\author{Tuo Zhang}
\affiliation{
  \department{Department of Computer Science}
  \institution{University of Birmingham}
  \city{Birmingham}
  \country{United Kingdom}}
\email{t.zhang.8@bham.ac.uk}
\orcid{0009-0000-2655-6486}

\author{Leonardo Stella}
\affiliation{
  \department{Department of Computer Science}
  \institution{University of Birmingham}
  \city{Birmingham}
  \country{United Kingdom}}
\email{l.stella@bham.ac.uk}
\orcid{0000-0002-2670-9873}

\begin{abstract}
Zero-sum games are a fundamental setting for adversarial training and decision-making in multi-agent learning (MAL). Existing methods often ensure convergence to (approximate) Nash equilibria by introducing a form of regularization. Yet, regularization requires additional hyperparameters, which must be carefully tuned--a challenging task when the payoff structure is known, and considerably harder when the structure is unknown or subject to change. Motivated by this problem, we repurpose a classical model in evolutionary game theory, i.e., the Brown-von Neumann-Nash (BNN) dynamics, by leveraging the intrinsic convergence of this dynamics in zero-sum games without regularization, and provide last-iterate convergence guarantees in noisy normal-form games (NFGs). Importantly, to make this approach more applicable, we develop a novel framework with theoretical guarantees that integrates the BNN dynamics in extensive-form games (EFGs) through counterfactual weighting. Furthermore, we implement an algorithm that instantiates our framework with neural function approximation, enabling scalable learning in both NFGs and EFGs. Empirical results show that our method quickly adapts to nonstationarities, outperforming the state-of-the-art regularization-based approach.
\end{abstract}

\keywords{Multi-agent learning; evolutionary dynamics; game theory; last-iterate convergence}

\newcommand{\BibTeX}{\rm B\kern-.05em{\sc i\kern-.025em b}\kern-.08em\TeX}

\begin{document}


\pagestyle{fancy}
\fancyhead{}


\maketitle 

\section{Intoduction}
With the pervasive advent of generative models and agentic AI, the research community has turned towards multi-agent learning (MAL) to lay the foundations for these multi-agent systems~\cite{tran2501multi}, despite the growing criticism on their intrinsic nature~\cite{botti2025agentic}. Since Shoham, Powers and Grenager initiated a vibrant debate across AI, economics, and engineering with their famous inquiry ``If multi-agent learning is the answer then what is the question?''~\cite{shoham2007if}, MAL has largely trailed behind game theory since its inception, with early breakthroughs such as Brown and Robinson’s fictitious play~\cite{brown1951iterative, robinson1951iterative}, designed to computationally validate von Neumann’s seminal results on mixed Nash equilibria in zero-sum games~\cite{von2007theory}. Yet, it is precisely in the high-stakes setting of learning zero-sum games with noisy feedback where obtaining theoretical guarantees of convergence to the Nash equilibria becomes crucial.

Existing approaches that provide such guarantees generally fall into two categories: time-average convergence, which ensures that the running average of strategies converges, and last-iterate convergence, a stronger form of convergence where the strategies themselves converge. Between these two notions of convergence, the former is often unsatisfactory in practice, and computationally costly in large-scale games.
Moreover, in many realistic settings, an endogenous environmental nonstationarity is added to the intrinsic nonstationarity resulting from the agents’ interactions~\cite{zhang2023multi,han2022learning}. In such cases, time averaging becomes ineffective as outdated historical information continues to influence the learning process. On the contrary, approaches that achieve last-iterate convergence provide a more stable and responsive notion of learning in dynamic environments. Existing approaches of this kind mainly rely on regularization or payoff perturbation, which stabilizes the learning dynamics through a policy-based regularization term to the objective. However, this additional term creates another problem, namely, the tuning of additional hyperparameters, which is challenging when the payoff structure is known, and becomes considerably harder when the environment is unknown or changes over time.

\begin{figure*}[t]
    \centering
    \begin{subfigure}[t]{0.4\textwidth}
        \centering
        \includegraphics[width=\linewidth]{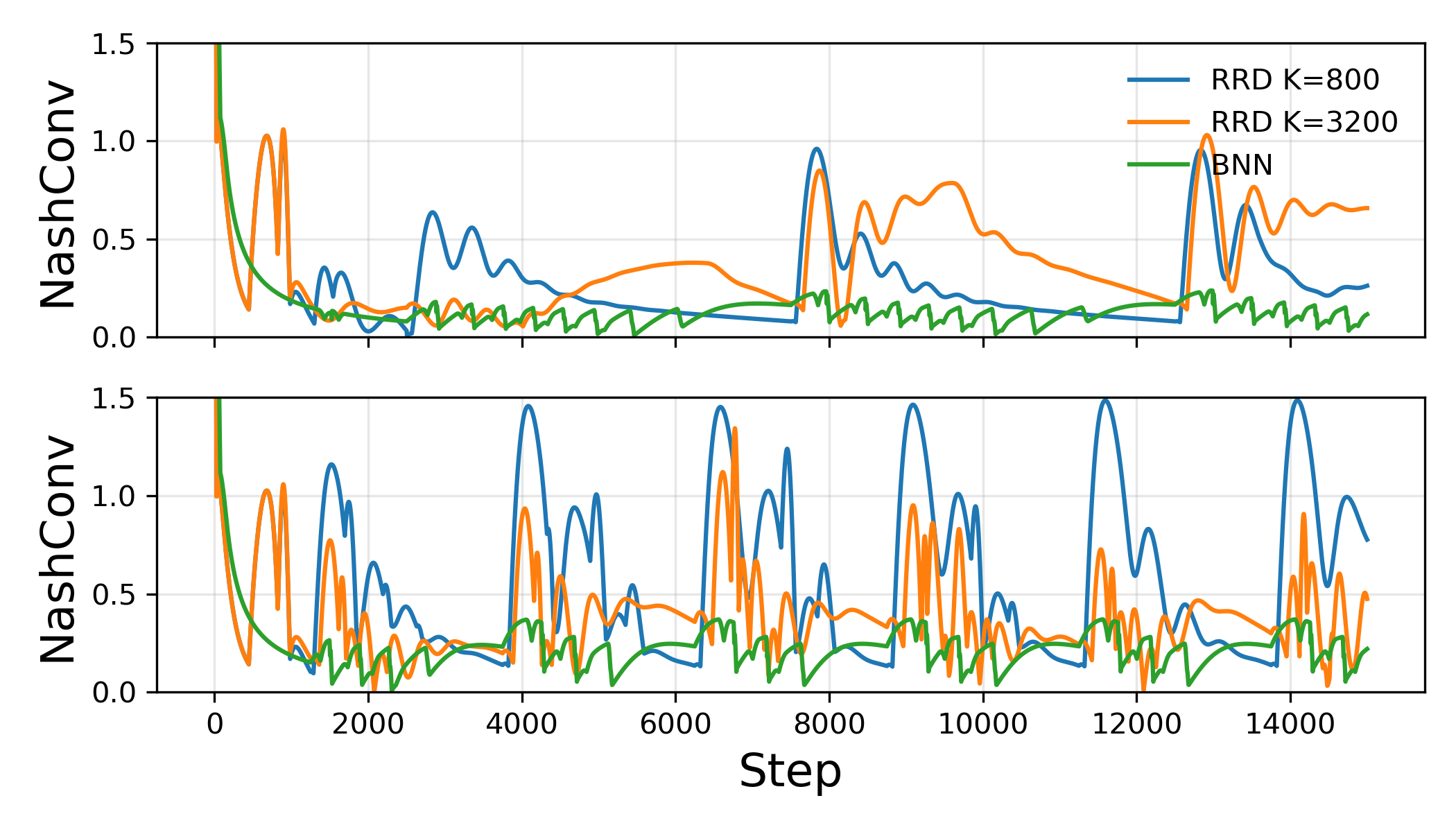}\label{fig:1a}
        \caption{\textsc{NashConv} metric in the nonstationary RPS.}
        \label{fig:left_panel}
    \end{subfigure}
    \hspace{6mm}
    \begin{subfigure}[t]{0.5\textwidth}
        \centering
        \includegraphics[width=\linewidth]{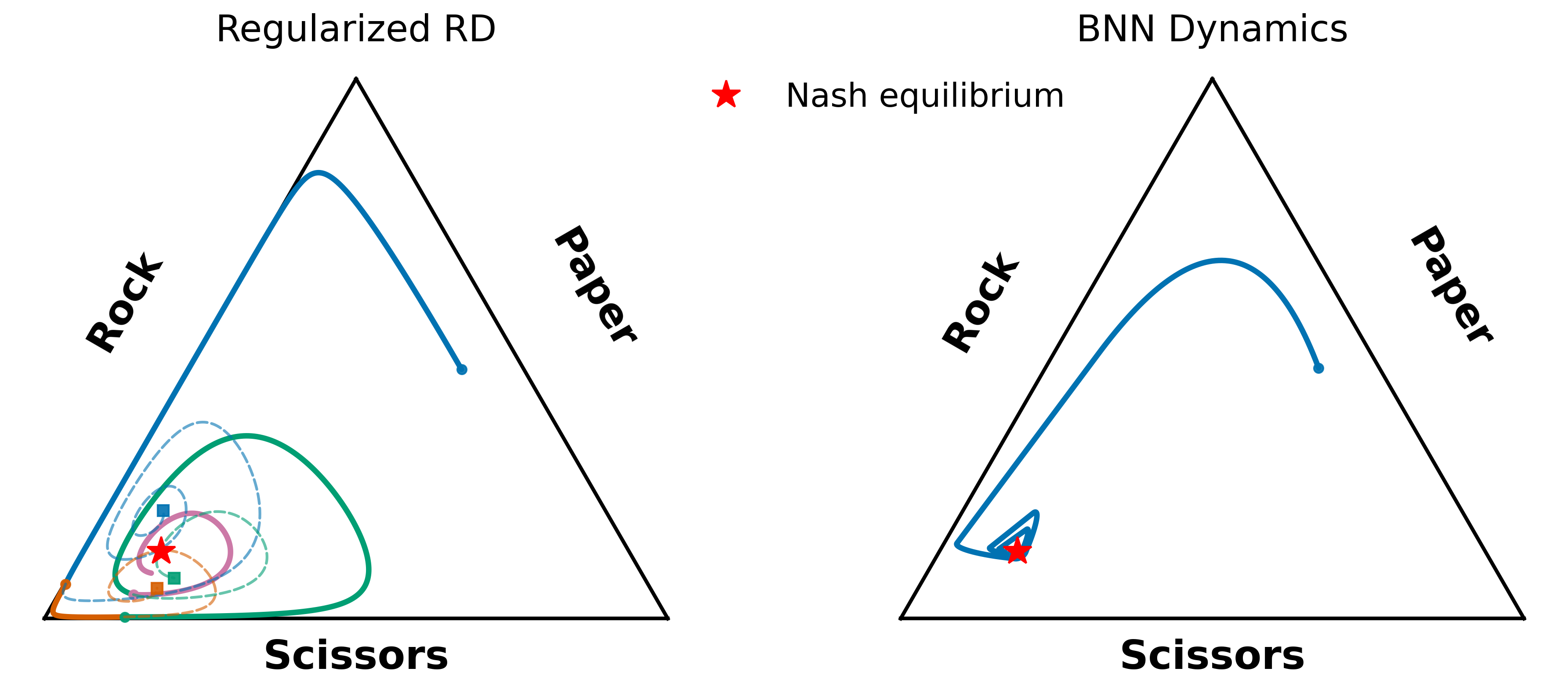}
        \caption{Representative trajectories and their convergence behavior.}
        \label{fig:right_panel}
    \end{subfigure}
    \caption{Comparison between the regularized RD and the BNN dynamics across different settings. \textsc{NashConv} metric in the nonstationary RPS environment with continuous changes in the payoffs (left). Representative trajectories under biased stationary payoffs in the simplex of the biased rock-paper-scissor (RPS) game (right).}
    \label{fig:motivation}
\end{figure*}


To further explain the issue caused by regularization, consider the following motivating example. The nonstationary Rock-Paper-Scissors (RPS) game is like the traditional RPS game but pairs of payoffs change continuously over time (within a time window of 2500 or 1250 iterations), ensuring that the zero-sum property of the game holds true at all time. Figure~\ref{fig:motivation}(a) shows the evolution of \textsc{NashConv}--a metric that captures the distance of the learning dynamics of an algorithm from the Nash equilibrium. In particular, we compare the Brown–von Neumann–Nash (BNN) dynamics, which our proposed approach is based on, with regularized replicator dynamics (RD), which forms the backbone of the state-of-the-art regularized Nash dynamics (R-NaD) algorithm~\cite{perolat2021poincare}. The BNN dynamics is a well-studied dynamical system in evolutionary game theory (EGT), with its key properties established in the classical literature (see, e.g., \cite{hofbauer2009brown,hofbauer2011deterministic}), but this is the first time it is introduced into MAL. Although both sets of dynamics retain the last-iterate convergence guarantee, this experiment shows the superiority of the BNN dynamics, in terms of proximity and stability around the Nash equilibrium against large oscillations of the regularized RD. 

This difference can be explained by the underlying mechanisms of the two sets of dynamics, as depicted on the simplex of the biased RPS game in Figure~\ref{fig:motivation} (b). The structure of BNN dynamics inherently ensures convergence in negative semi-definite zero-sum games without requiring any additional regularization. In contrast, the regularized RD introduces a reference policy as a regularization term, effectively running the learning process on a perturbed zero-sum game whose payoffs are modified by the strength of the regularization. This modification shifts the system’s attractor to a new point, which is determined jointly by the original Nash equilibrium, the reference policy, and the regularization coefficient. By periodically updating the reference policy or decreasing the regularization strength, the new attractor gradually approaches the true Nash equilibrium. Consequently, the switching frequency and the regularization strength become two crucial hyperparameters. 
When the payoffs themselves change, this mechanism can become unsafe, as the coupling between the regularization dynamics and the evolving environment may drive the system towards unpredictable or undesired behaviors. To better visualize the described behavior, trajectories of the regularized RD are color-coded to indicate different reference stages of the process and different attractors. The BNN dynamics naturally converges to the Nash equilibrium without the need for regularization. 

Motivated by this critical limitation in the state-of-the-art regularized RD methods, we propose the following contribution.
\begin{enumerate}
\item We establish a theoretical MAL framework grounded in BNN dynamics for two-player zero-sum games with noisy feedback, both in normal-form and in extensive-form settings. 
\item Building on this framework, we develop an algorithm based on an actor–critic (AC) architecture that implements the population-level BNN dynamics. 
\item We validate our algorithm on a comprehensive benchmark of zero-sum games, including both normal-form, extensive-form, and nonstationary settings. Our experiments show that its empirical behavior aligns with our theoretical predictions and that the algorithm outperforms state-of-the-art baselines based on regularized RD methods, showing superiority of our approach in terms of stability, convergence, and safety.
\end{enumerate}

\subsection{Related Work}
Early connections between EGT and MAL were established by Börgers and Sarin~\cite{borgers1997learning}, who provided the first formal link between the RD in EGT and the cross learning algorithm~\cite{cross1973stochastic}, later 
inspiring extensions to more sophisticated methods such as Q-learning and its variants~\cite{tuyls2003selection,kianercy2012dynamics,abdallah2008multiagent}, as well as regret-minimization approaches~\cite{kaisers2012common}. Subsequent work modified learning rules so that they inherit stability properties from evolutionary processes~\cite{kaisers2010frequency,perolat2021poincare}, and extended the framework to a broader range of games and dynamics~\cite{vrancx2008switching,hennes2009state,hennes2010resq,mertikopoulos2018riemannian}. Recently, EGT has also been used to provide theoretical grounding for algorithms with function approximation and deep learning~\cite{hennes2020neural,perolat2022mastering,barfuss2019deterministic,lanctot2017unified}. In our recent work, we have explored innovative dynamics, as opposed to imitation-based rules such as RD~\cite{zhang2025experience}.

Parallel to the evolutionary approach, the line of work based on regularization originates from the idea of payoff perturbation~\cite{facchinei2003finite}. By introducing a strongly convex penalty to the players’ payoff or utility functions, these methods ensure convergence towards approximate Nash equilibria~\cite{cen2021fast,cen2022faster,pattathil2023symmetric}. For example, Sokota \emph{et al.}~\cite{sokota2022unified} demonstrated that their perturbed mirror descent algorithm converges to the quantal response equilibrium~\cite{mckelvey1995quantal,mckelvey1998quantal}. In continuous-time settings, numerous studies have also reported similar results~\cite{coucheney2015penalty,leslie2005individual,abe2022mutation,hussain2023asymptotic}. However, ensuring convergence requires careful adjustment of the perturbation strength~\cite{liu2022power,cai2023uncoupled}, with related studies exploring iterative Tikhonov regularization schemes~\cite{koshal2010single,tatarenko2019learning,yousefian2017smoothing}. Perolat \emph{et al.}~\cite{perolat2021poincare} introduced the idea of payoff perturbations that depend on a reference (or regularized) strategy, periodically updated to promote convergence. Later studies refined the switching mechanism to achieve faster convergence~\cite{abe2023adaptively,abe2024boosting,abe2022last}.

Another prominent class of methods to achieve last-iterate convergence relies on the idea of \emph{optimistic learning}~\cite{rakhlin2013online,rakhlin2013optimization}. Examples include the Optimistic Follow the Regularized Leader~\cite{shalev2006convex}, Optimistic Mirror Descent algorithms~\cite{zhou2017mirror,hsieh2021adaptive}, and more recently Smooth Optimistic Gradient-Based Regret Matching+~\cite{meng2025sogrm}. Such approaches have been shown to attain last-iterate convergence across a wide range of games~\cite{daskalakis2017training,daskalakis2018last,daskalakis2018limit,mertikopoulos2018optimistic}, with established convergence rates~\cite{gorbunov2022last,cai2023doubly}. However, these results typically assume accurate and noiseless gradient feedback, and their performance deteriorates in the presence of stochastic or biased observations~\cite{wei2020linear,lee2021fast}. In some cases, negative empirical results have also been reported~\cite{abe2023adaptively}.

\section{Preliminaries}\label{sec:background}
This section provides the background and notation used throughout the paper. The remainder of this work is based on the concepts and formulations introduced here.

\subsection{Normal-form Games}

Normal-form games provide the static, population-level foundation for analysing learning dynamics in multi-agent systems. 
We consider two-player zero-sum games with finite action sets $A_i$ 
and utility functions $u_i : A_1 \times A_2 \to [-u_{\max},\,u_{\max}]
$ for each player $i\in\{1,2\}$,
satisfying $u_1(a_1,a_2)=-u_2(a_1,a_2)$ for all $(a_1,a_2)\in A_1\times A_2$. A mixed strategy of player $i$ is a probability distribution $\pi_i \in \Delta(A_i)$ over its actions, 
where $\pi_i(a_j)$ denotes the probability of selecting action $a_j \in A_i$ under $\pi_i$.

The expected payoff under a strategy profile $\pi=(\pi_1,\pi_2)$ is 
$$
u_i(\pi_1,\pi_2)
 = \sum_{a_j\in A_1}\sum_{a_k\in A_2}
   \pi_1(a_j)\pi_2(a_k)\,u_i(a_j,a_k).
$$

A strategy profile $\pi^*=(\pi_1^*,\pi_2^*)$ is a \textbf{Nash equilibrium} if, for each player $i$, 
$$
u_i(\pi_i^*,\pi_{-i}^*) \ge u_i(\pi_i,\pi_{-i}^*), \quad \forall\,\pi_i\in\Delta(A_i).
$$

For each pure action $a_j \in A_i$, we define its expected payoff against the opponent’s strategy as
$$
u_i(a_j \mid \pi_{-i}) 
= \sum_{a_k \in A_{-i}} \pi_{-i}(a_k)\,u_i(a_j,a_k).
$$
For compactness, we write $u_i(a_j)$ to denote $u_i(a_j \mid \pi_{-i})$, 
and use $\bar u_i = u_i(\pi_1,\pi_2)$ to denote the expected payoff under the mixed strategy profile~$\pi$.





\subsection{Extensive-form Games}
\label{sec:efg}

We now extend the normal-form notation introduced above to sequential games with imperfect information.  
An extensive-form game for two players is represented as a tuple
\[
G = (H, Z, (A(h))_{h \in H \setminus Z}, (\tau(h))_{h \in H \setminus Z}, (u_i)_{i=1,2}, (\mathcal{I}_i)_{i=1,2}),
\]
where 
$H$ is the finite set of all possible histories (action sequences), 
and $Z \subset H$ is the set of terminal histories.  
At each non-terminal history $h \in H \setminus Z$, 
player $\tau(h) \in \{1,2\}$ chooses an action from the available set $A(h)$.  
Histories that player~$i$ cannot distinguish are grouped into information sets 
$I \in \mathcal{I}_i$, where $\mathcal{I}_i$ denotes the collection of all such sets for player~$i$.

A \emph{behaviour strategy} of player~$i$ is a mapping 
$\sigma_i : \mathcal{I}_i \to \Delta(A(I))$, 
assigning a probability distribution over available actions at each information set~$I$.  
The joint strategy profile is written $\sigma = (\sigma_1, \sigma_2)$.

\paragraph{Reach probabilities.}
Given $\sigma$, the probability of reaching a history $h=(a_1,\ldots,a_t)$ is
\begin{equation}
\rho_\sigma(h)
=\prod_{k=1}^{t}
   \sigma_{\tau(h_{k-1})}\!\big(a_k \mid I(h_{k-1})\big),
\end{equation}
which can be factorised as $\rho_\sigma(h)=\rho^i_\sigma(h)\rho^{-i}_\sigma(h)$, 
representing the contributions of player~$i$ and its opponents, respectively.  
The \emph{reach probability} of an information set $I \in \mathcal{I}_i$ is then
\begin{equation}
\rho_\sigma(I)=\sum_{h\in I}\rho_\sigma(h),
\qquad
\rho^{-i}_\sigma(I)=\sum_{h\in I}\rho^{-i}_\sigma(h),
\end{equation}
and under perfect recall we have $\rho_\sigma(I)=\rho^i_\sigma(I)\rho^{-i}_\sigma(I)$ for all $I$.

\paragraph{Counterfactual values.}
For each $I\in\mathcal{I}_i$ and action $a\in A(I)$, 
the \emph{counterfactual value} of player~$i$ is
\begin{equation}
v_i^\sigma(I,a)
=\sum_{h\in I}\rho^{-i}_\sigma(h)
   \sum_{\substack{z\in Z\\ h a \prec z}}
   \rho_\sigma(z\mid h a)\,u_i(z),
\end{equation}
where $u_i(z)$ is the terminal payoff 
and $h a \prec z$ indicates that terminal history $z$ extends the partial history $h a$.  
This expression isolates the expected return from taking $a$ at $I$, 
weighted by the opponent’s reach probability $\rho^{-i}_\sigma(h)$, 
and thus corresponds to the \emph{counterfactual Q-function} in extensive-form games.  
The expected value at $I$ is then
\begin{equation}
v_i^\sigma(I)=\sum_{a} \sigma_i(a|I)\,v_i^\sigma(I,a).
\end{equation}

A strategy profile $\sigma^*=(\sigma_1^*,\sigma_2^*)$ is a \emph{Nash equilibrium} 
if no player can unilaterally improve its expected payoff:
\begin{equation}
V_i(\sigma_i',\sigma_{-i}^*) \le V_i(\sigma^*) ,\qquad \forall\sigma_i'.
\end{equation}

\subsection{BNN Dynamics}
The BNN dynamics belongs to the family of dynamics called innovative dynamics, as new strategies can enter the population through experimentation, exploration and mutation, and describes the continuous-time evolution of strategies. For player $i$ and action $a \in A_i$, the BNN dynamics takes the form:
$$
\dot{\pi}_i(a) = [\,u_i(a) - \bar u_i\,]_+ \;-\; \pi_i(a)\sum_{a' \in A_i}[\,u_i(a') - \bar u_i\,]_+ .
$$  

These dynamics preserve the strategy simplex and guarantee no regret in the limit.  

A discrete-time stochastic approximation with step size $\eta_t>0$ takes the form  
$$
\pi_i^{t+1}(a) = \pi_i^t(a) + \eta_t \Big( [\,u_i(a) - \bar u_i\,]_+ - \pi_i^t(a)\sum_{a' \in A_i}[\,u_i(a') - \bar u_i\,]_+ \Big).
$$

\subsection{Noisy Feedback}

In practice, players often rely on estimated utilities rather than exact payoffs, due to sampling or inherent stochasticity of the environment. To capture this, we model feedback as a noisy observation. For each action $a \in A_i$, the observed payoff is  
$$\tilde u_i(a) = u_i(a) + \xi_i(a),$$  
and the corresponding average payoff under profile $\pi$ is  
$$\tilde{\bar u}_i = \sum_{a_i \in A_i} \pi_i(a_i)\,\tilde u_i(a_i).$$  

The noise $\xi_i(a)$ represents estimation errors or random perturbations. Following standard assumptions in stochastic approximation and learning in games, we require the noise to be unbiased with uniformly bounded variance.

\begin{assumption}\label{ass:noisy}
For each player $i$ and action $a \in A_i$, the noise satisfies  
$$\mathbb{E}[\xi_i(a)] = 0, \qquad \mathrm{Var}(\xi_i(a)) \le \sigma^2 < \infty.$$  
\end{assumption}

This ensures that observed payoffs remain correct on average, while the variance bound $\sigma^2$ provides a uniform control of uncertainty and will enter directly into our theoretical results.

\section{BNN Dynamics under Noisy Feedback}\label{sec:theory}

We now analyze the BNN dynamics under noisy feedback in the normal-form setting.  Our aim is to characterize how stochastic perturbations influence the learning trajectory and whether the convergence guarantees of the deterministic process persist in this stochastic regime. This section provides a foundation for examining stability and bias before extending the analysis 
to more general sequential games.

\subsection{Discrete Update and SA Form}

Having established the noiseless BNN dynamics in Section~2.3, we now turn to the setting with noisy feedback. To analyse the resulting stochastic recursion, it is convenient to introduce the vector-field notation. Let $H(\pi)$ denote the noiseless update direction defined in Section~2.3. Under noisy feedback (Assumption~\ref{ass:noisy}), the recursion becomes
$$
\pi_{t+1} = \pi_t + \eta_t \,\widehat H(\pi_t),
$$
with noisy field components
$$
\widehat H_i(a;\pi) = [\,\tilde u_i(a)-\tilde{\bar u}_i\,]^+ - \pi_i(a)\sum_{a'\in A_i}[\,\tilde u_i(a')-\tilde{\bar u}_i\,]^+ .
$$

Because $[\,\cdot\,]_+$ is convex, the expectation of the noisy update differs from the noiseless field, creating a structural bias (a Jensen gap). Formally, the noisy field admits the decomposition
$$
\mathbb{E}[\widehat H(\pi_t)] = H(\pi_t) + \beta(\pi_t),
$$
where $\beta(\pi)$ denotes the bias.

To proceed with the stochastic approximation analysis, it is important to understand the size of this bias.  

\begin{lemma}\label{bias_bound}
    
For a player with action set $A_i$, the structural bias term $\beta(\pi)$ is uniformly bounded over the simplex as
$$
\|\beta(\pi)\|_{\infty} \le (|A_i|-1)\,\sigma,
$$
where $\sigma^2$ is the variance bound in Assumption~\ref{ass:noisy}.
\end{lemma} 


In addition to the bias, the decomposition also produces a residual noise term. Define
$$
\zeta_{t+1} := \widehat H(\pi_t) - \mathbb{E}[\widehat H(\pi_t)] .
$$
This is a martingale-difference sequence with zero conditional mean, i.e.\ $\mathbb{E}[\zeta_{t+1}\mid \pi_t] = 0$. 
Since $[\,\cdot\,]_+$ is 1-Lipschitz and the payoff noise has bounded variance by Assumption~\ref{ass:noisy}, the variance of $\zeta_{t+1}$ is uniformly bounded. In particular,
$$
\mathbb{E}[\|\zeta_{t+1}\|^2 \mid \pi_t] \le (1+|A_i|)^2\sigma^2 .
$$
For brevity we henceforth write this bound as $ C\sigma^2$ .

Collecting the bias and noise terms, the recursion takes the stochastic approximation form
$$
\pi_{t+1} = \pi_t + \eta_t \big( H(\pi_t) + \beta(\pi_t) + \zeta_{t+1} \big).
$$

\subsection{Convergence Results}\label{sec:convergence}

To analyse convergence, we first recall that the continuous-time BNN flow $\dot\pi = H(\pi)$ admits a Lyapunov potential. Specifically,
$$
\Gamma(\pi) := \tfrac12 \sum_{a\in A_i} \bigl[u_i(a;\pi)-u_i(\pi)\bigr]_+^2
$$
serves as a strict Lyapunov function in normal-form zero-sum games (see, e.g., [ref]). 
For convenience, we also introduce
$$
S(\pi) := \sum_{a\in A_i} \bigl[u_i(a;\pi)-u_i(\pi)\bigr]_+ ,
$$
which represents the aggregate mass of profitable deviations. 
The two quantities are linked by the elementary inequality
$S(\pi)^2 \;\ge\; 2\,\Gamma(\pi)$.

For the noiseless BNN flow, the Lyapunov potential satisfies the differential identity
$$
\frac{d}{dt}\Gamma(\pi) = -2\,S(\pi)\,\Gamma(\pi).
$$
Together with the inequality $S(\pi)^2 \ge 2\Gamma(\pi)$ established above, this implies
$$
\frac{d}{dt}\Gamma(\pi) \;\le\; -2\sqrt{2}\,\Gamma(\pi)^{3/2},
$$
demonstrating strict dissipation whenever profitable deviations remain.

In the presence of noisy feedback, the exact dissipation identity no longer holds. The structural bias $\beta(\pi)$ perturbs the dynamics and the martingale noise adds further fluctuations, so the Lyapunov descent is weakened. In the discrete recursion this effect is captured by a one-step drift inequality in expectation, which combines the dissipative behaviour of the noiseless flow with the perturbations due to bias and noise.

Combining the bias bound of Lemma~\ref{bias_bound} with the bounded-variance noise, 
and under Robbins--Monro step size conditions, we obtain the following drift inequality.  

\begin{lemma}\label{lem:expected-descent-Gamma}  
Denote $g_t := \mathbb{E}[\Gamma(\pi_t)]$. 
Suppose the step sizes $\{\eta_t\}$ satisfy $\sum_t \eta_t = \infty$ and $\sum_t \eta_t^2 < \infty$ 
(e.g.\ $\eta_t = c/(t+t_0)^{2/3}$). Then
$$
g_{t+1} \;\le\; g_t \;-\; 2\sqrt{2}\,\eta_t\, g_t^{3/2} 
\;+\; (|A_i|-1)\sqrt{2|A_i|}\,\sigma\,\eta_t\, g_t^{1/2} 
\;+\; C_3\,\eta_t^2 ,
$$
where $C_3 > 0$ is a finite constant depending only on the game, 
but independent of $\sigma$ and $t$.
\end{lemma}

The constant $C_3$ arises from the quadratic remainder of the Taylor expansion. 
It depends on game-specific Lipschitz constants and a uniform bound on the second moments of the noise. 
Since it always appears multiplied by $\eta_t^2$ and $\sum_t \eta_t^2 < \infty$, 
this contribution is summable and does not affect the asymptotic analysis.

The one-step inequality of Lemma~\ref{lem:expected-descent-Gamma} shows that the Lyapunov potential decreases in expectation, up to two perturbations: the quadratic remainder $C_3\eta_t^2$ and the linear bias term proportional to $\sigma$. Building on this descent, we now establish two global consequences: (i) almost sure stability within an $O(\sigma)$ neighbourhood, and (ii) a quantitative rate of approach to this noise floor. Under the standard Robbins--Monro step-size conditions for stochastic approximation, we obtain the following stability bound. These conditions ensure that the iterates asymptotically follow the mean-field dynamics while the accumulated variance remains finite.

\begin{theorem}\label{thm:stability}
Suppose the step sizes $\{\eta_t\}$ satisfy $\sum_t \eta_t = \infty$ and 
$\sum_t \eta_t^2 < \infty$. Then
$$
\limsup_{t\to\infty} \Gamma(\pi_t) \;\le\; \tfrac12 (|A_i|-1)\sqrt{|A_i|}\,\sigma \quad \text{almost surely}.
$$
\end{theorem}

This theorem shows that the iterates stabilise almost surely within an $O(\sigma)$ neighbourhood of the Nash set, so the asymptotic error floor scales linearly with the noise level. The proof relies on completing the square in the drift inequality and applying the Robbins--Siegmund supermartingale theorem; full details are provided in Appendix.  

While Theorem~\ref{thm:stability} establishes asymptotic stability, it does not quantify the speed of convergence. To address this, we analyse the transient decay of $\Gamma$ until it reaches the $O(\sigma)$ floor. This is achieved by rewriting the drift inequality as a discrete Riccati-type recursion and comparing it with an auxiliary differential equation.  

\begin{theorem}\label{thm:rate}
Let $\eta_t = c/(t+t_0)^{2/3}$ with $c>0$ and $t_0 \ge 1$. Then
$$
\mathbb{E}[\Gamma(\pi_t)] = O(t^{-2/3})
$$
until it stabilises at the $O(\sigma)$ floor described in Theorem~\ref{thm:stability}. 
In the noiseless case $(\sigma=0)$, the rate $O(t^{-2/3})$ holds for all $t$.
\end{theorem}

Theorem~\ref{thm:rate} shows that the decay exponent $2/3$ coincides with the noiseless case; noise only determines the eventual $O(\sigma)$ floor. The proof is based on a discrete comparison argument using the substitution $y_t = \sqrt{\mathbb{E}[\Gamma(\pi_t)]}$; details are provided in Appendix. Two further remarks are worth noting:  
(i) with a constant step size $\eta$, the process attains a steady-state error of order $O(\sigma\eta) + O(\eta^2)$;  
(ii) the convergence is not in a standard norm but in a payoff-based measure, which coincides with exploitability and regret. This perspective makes the result more widely applicable, as elaborated in the next section.

\subsection{Bias and Centroid Shift}

Theorem~\ref{thm:stability} established that the noisy recursion stabilises within an $O(\sigma)$ neighbourhood in $\Gamma$, reflecting the stochastic fluctuations of the process. We now turn to a different question: how far the stationary points of the biased dynamics themselves are displaced from Nash equilibria. This effect is deterministic, arising solely from the structural bias $\beta$, and requires a separate line of analysis based on bounding the maximal advantage at a biased fixed point.  

Formally, a biased fixed point $\pi'$ is defined by  
$$
H(\pi') + \beta(\pi') = 0.
$$  
At such a point, the maximal advantage satisfies $\max_{a\in A_i} \text{Adv}(a) \le (|A_i|-1)\sigma$. Since the Lyapunov potential is quadratic in advantages, this yields the following bound.  

\begin{theorem}[Centroid shift]\label{thm:centroid}
Let $\pi'$ be a fixed point of the biased recursion, i.e.\ $H(\pi')+\beta(\pi')=0$. Then
$$
\Gamma(\pi') \;\le\; \tfrac12 |A_i| (|A_i|-1)^2 \sigma^2 .
$$
\end{theorem}

This shows that structural bias shifts the centroid of the dynamics only quadratically in $\sigma$: biased fixed points lie within an $O(\sigma^2)$ neighbourhood in $\Gamma$. In contrast, Theorem~\ref{thm:stability} describes a stochastic noise floor of order $O(\sigma)$. Taken together, these results imply that as the variance of the noise vanishes ($\sigma^2 \to 0$), both the fluctuation floor and the centroid shift disappear, and the iterates converge to exact Nash equilibria.

\section{BNN Dynamics in Extensive-Form Games}
Having established the properties of the BNN dynamics in normal-form games, we now extend the framework to extensive-form games with imperfect information. This setting captures sequential decision processes and therefore makes our framework more relevant to practical multi-agent learning and decision-making scenarios. The following section reformulates the instantaneous regret descent in terms of local information sets and introduces reach-weighted updates that generalize the normal-form dynamics to sequential play.

\subsection{Local Regret Structure and Reach Weights}

In normal-form games the dynamics are driven by the positive part of the payoff difference, $[\Delta_i(a\mid\pi)]_+$, which is exactly the instantaneous regret of action $a$. The aggregate $S_i(\pi)=\sum_a [\Delta_i(a\mid\pi)]_+$ is the total regret mass considered in regret--minimisation. Moreover, $S_i(\pi)$ is bounded (up to a normalisation) by the potential $\Gamma(\pi)$ introduced in Section~3, which is non--increasing along the flow and functions as a Lyapunov function for the dynamics. Because $S_i$ is the sum of positive regrets, this identification provides the minimal bridge from the NFG analysis to the extensive-form case.

In the extensive-form setting the analogue of this signal is defined at each information set. For player $i$ at information set $x$, let $u_i^\pi(x,a)$ be the expected payoff conditional on reaching $x$ and choosing action $a$, and let $\bar u_i^\pi(x)=\langle u_i^\pi(x,\cdot),\pi_i(\cdot\mid x)\rangle$. The difference
\[
A_i^\pi(x,a)=u_i^\pi(x,a)-\bar u_i^\pi(x)
\]
is then used in place of $\Delta_i(a\mid\pi)$. Its positive part coincides with the local regret notion at information sets, making it a natural replacement for the normal-form signal.

Using only the payoff differences at information sets would still fail to capture their correct impact on the overall dynamics. A local deviation at $x$ contributes to the game outcome only in proportion to the probability that the play actually reaches $x$, which is determined by the opponent’s reach probability $\rho_{-i}^\pi(x)$. To obtain the correct scaling, the update at each information set must therefore be multiplied by this factor. The resulting extensive-form dynamics take the form  
\[
\dot \pi_i(a\mid x) \;=\; \rho_{-i}^\pi(x)\Big( [A_i^\pi(x,a)]_+ \;-\; \pi_i(a\mid x)\sum_{b\in A(x)} [A_i^\pi(x,b)]_+ \Big),
\]  
which is the direct analogue of the BNN dynamics in normal-form games, now lifted to the information-set decomposition with reach weighting.

\subsection{Theoretical Results in EFG}

We now turn to the theoretical analysis of the extensive-form dynamics introduced above. All results from Section 3 extend directly once local counterfactual advantages are weighted by the opponent’s reach probabilities. To avoid repetition, we state the reach-weighted counterparts of the normal-form results and highlight their parallels.

We first focus on the noiseless continuous-time dynamics. 
Define the reach-weighted potential
\[
V(\pi) = \sum_{i}\sum_{x\in\mathcal{X}_i}\Gamma^x_i(\pi),
\qquad 
\Gamma^x_i(\pi) = \tfrac12 \big\|\rho^\pi_{-i}(x)\,A^\pi_{i,+}(x,\cdot)\big\|^2 .
\]

\begin{lemma}
Along the noiseless extensive-form dynamics, the potential $V(\pi)$ satisfies the exact identity
\begin{align*}
\frac{d}{dt}V(\pi) 
&= - \sum_{i}\sum_{x\in\mathcal{X}_i} 2\,S^x_i(\pi)\,\Gamma^x_i(\pi),
\\
\qquad 
S^x_i(\pi) 
&= \sum_{a\in A(x)} \rho^\pi_{-i}(x)\,[A^\pi_i(x,a)]_+ .
\end{align*}
In particular, $\dot V(\pi)<0$ whenever $\pi\notin\Pi^\star$, where $\Pi^\star$ denotes the equilibrium set.
\end{lemma}

Because BNN dynamics preserve the interior of the simplex, all actions maintain strictly positive probability, so every relevant reach probability $\rho^\pi_{-i}(x)$ remains bounded away from zero. This ensures that the weighted dissipation in the identity is strictly positive outside equilibrium, establishing $V$ as a Lyapunov function.

We now turn to the stochastic recursion with noisy counterfactual feedback. 
To distinguish from the normal-form setting, we use a prime notation to denote the extensive-form counterparts of the quantities in Section~3, e.g.,\ $H'$, $\beta'$, and $\zeta'$.

With Robbins--Monro step sizes $\{\alpha_k\}$, the update at each information set $x \in \mathcal{X}_i$ takes the form
\begin{align*}
\pi^{k+1}_i(a|x) 
&= \pi^k_i(a|x) + \alpha_k \,\rho^{\pi^k}_{-i}(x)\\
&\Big([\,\widetilde A^{\pi^k}_i(x,a)\,]_+ - \pi^k_i(a|x)\!\!\sum_{b\in A(x)} [\,\widetilde A^{\pi^k}_i(x,b)\,]_+\Big),
\end{align*}
where $\widetilde A^{\pi^k}_i(x,a)$ is a possibly noisy estimator of the counterfactual advantage.

As in Section~3, the convexity of $[\cdot]_+$ creates a structural bias $\beta'(\pi)$, while the sampling error yields a martingale-difference noise $\zeta'_{k+1}$ with bounded variance. 
The form of these bounds is identical to the normal-form case, but the constants now depend on the local action set sizes $\{|A(x)|:x\in\mathcal{X}_i\}$.

\begin{lemma}[Bias and noise bounds in EFG]
There exist finite constants $C_1,C_2 > 0$, depending only on the information-set structure, such that for all $\pi$
\[
\|\beta'(\pi)\| \;\le\; C_1\sigma,
\qquad
\mathbb{E}[\|\zeta'_{k+1}\|^2 \mid \pi^k] \;\le\; C_2\sigma^2,
\]
where $\sigma^2$ is the uniform variance bound of the payoff noise.
\end{lemma}

Collecting these terms, the recursion admits the stochastic approximation representation
\[
\pi^{k+1} = \pi^k + \alpha_k\big(H'(\pi^k) + \beta'(\pi^k) + \zeta'_{k+1}\big),
\]
where $H'$ denotes the noiseless extensive-form BNN field.

With these bounds in place, the Lyapunov potential admits a quantitative descent estimate in expectation, stated as follows.

\begin{lemma}[One-step expected descent in EFG]
Let $g_k := \mathbb{E}[V(\pi^k)]$, where $V$ is the Lyapunov function defined in Lemma~4.2.1. 
Suppose the step sizes $\{\alpha_k\}$ satisfy $\sum_k \alpha_k = \infty$ and $\sum_k \alpha_k^2 < \infty$. 
Then there exist constants $C_1,C_2 > 0$, depending only on the information-set structure, such that
\[
g_{k+1} \;\le\; g_k - 2\alpha_k \rho \, g_k^{3/2} + C_1 \sigma \alpha_k g_k^{1/2} + C_2 \alpha_k^2,
\]
where $\rho>0$ is a uniform lower bound on reach probabilities in the interior of the simplex, 
and $\sigma^2$ is the variance bound of the payoff noise.
\end{lemma}

Building on the one-step descent inequality, we now establish the asymptotic behaviour of the stochastic recursion in the extensive-form setting. 
The statements mirror those of Section~3, but with constants depending only on the information-set structure. 
For completeness, we collect the results here using the same Lyapunov potential $V(\pi)$.

\begin{theorem}[Asymptotic stability in EFG]
Let $\{\pi^k\}$ follow the stochastic recursion defined above with Robbins--Monro step sizes. 
Then the Lyapunov potential satisfies
\[
\limsup_{k\to\infty}\,\mathbb{E}[V(\pi^k)] = O(\sigma),
\]
where $\sigma^2$ is the variance of the payoff noise.
\end{theorem}

\begin{theorem}[Convergence rate in EFG]
Suppose the step sizes are chosen as $\alpha_k = c/(k+t_0)^{2/3}$ with constants $c,t_0>0$. 
Then the expected potential decreases at rate
\[
\mathbb{E}[V(\pi^k)] = O(k^{-2/3}) + O(\sigma),
\]
so the iterates converge to an $O(\sigma)$ neighbourhood of equilibrium at this rate.
\end{theorem}

\begin{theorem}[Centroid shift in EFG]
In the presence of persistent noise, the stationary points of the recursion correspond to approximate equilibria where the Lyapunov potential satisfies
\[
V(\pi) = O(\sigma^2).
\]
Thus the asymptotic bias induced by noise manifests only at order $O(\sigma^2)$.
\end{theorem}

These results confirm that the last-iterate guarantees obtained in the normal-form analysis extend directly to extensive-form games once counterfactual advantages are employed. 
In particular, the asymptotic stability and convergence rate are limited also by an $O(\sigma)$ noise floor, while the systematic bias introduced by noise appears only at order $O(\sigma^2)$. 
Thus the extensive-form recursion inherits the same robustness properties as in the normal-form case, with no additional complications beyond the dependence of constants on the local action set sizes.

\subsection{Algorithmic Implementation}

We implement the BNN Actor–Critic (BNNAC) within an actor–critic framework. A key feature of our design is that all three components—the actor, critic, and reach estimator—are represented by neural networks. While a tabular form would be sufficient for small games, extensive–form games can be extremely large, and neural networks provide the scalable function approximation required to handle such settings efficiently. The algorithm is distributed: both players run BNNAC independently and simultaneously, without direct communication, and their interaction occurs only through the trajectories generated by the underlying game.

The actor-logits network $L_{\theta^L}$ outputs logits, which are mapped to a policy via a softmax transformation. The critic network $Q_{\theta^Q}$ predicts payoffs and is used to compute local advantage estimates. The reach network $\rho_{\theta^\rho}$ estimates opponent reach probabilities at information sets. Together these three networks provide the quantities required by the extensive-form BNN update.

Each training iteration obtains data $(x,a,u,x')$ from the game under the current policy. The critic parameters $\theta^Q$ are updated to reduce the prediction error of observed payoffs, while the reach parameters $\theta^\rho$ are adjusted according to the observed visitation frequencies. Every $K$ iterations, the actor is updated: advantages are computed from the critic, reach probabilities from the reach network, and the actor logits are then moved in the BNN direction before being mapped through the softmax transformation.

This design follows the theoretical analysis in two ways. First, the critic and reach networks are updated more frequently to ensure that the feedback received by the actor can be treated as approximately unbiased. Second, the softmax ensures that the policy remains on the simplex. Since the actor is parameterized in logits, propagating the update through softmax introduces an additional $\tfrac{1}{\pi}$ factor in the update direction. Importantly, BNN dynamics never eliminate strategies completely, so the denominator is always well defined; in practice, a small positive floor is used to prevent numerical underflow. Finally, introducing a separate reach network cleanly separates structural counterfactual weighting from payoff learning, which both clarifies the theoretical correspondence and stabilizes training. If the reach probabilities are omitted, the same structure directly reduces to the normal-form BNN actor–critic framework.

\begin{algorithm}[t]
\caption{BNN Actor--Critic (BNNAC)}
\label{alg:bnnac}
\begin{algorithmic}[1]
\State \textbf{Input:} Initial parameters for three networks: 
actor-logits $\theta^L_0$ (logit network $L_{\theta^L}$, policy $\pi=\mathrm{softmax}(L_{\theta^L})$), 
critic $\theta^Q_0$ ($Q_{\theta^Q}$ predicts payoffs), 
reach $\theta^\rho_0$ ($\rho_{\theta^\rho}$ estimates opponent reach). 
Step sizes $\{\eta_t\}, \{\alpha_t\}, \{\beta_t\}$ and update interval $K$.

\For{$t=0,1,2,\dots$}
  \State Obtain data $\mathcal{D} = \{(x,a,r,x')\}$ from the game

  \State Construct loss $\ell(\theta^Q_t;\mathcal{D})$ from the data
  \State Update critic: $\theta^Q_{t+1} \gets \theta^Q_t - \alpha_t \nabla_{\theta^Q} \ell(\theta^Q_t;\mathcal{D})$

  \State Update reach network: $\theta^\rho_{t+1} \gets \theta^\rho_t - \beta_t \nabla_{\theta^\rho} \ell_\rho(\theta^\rho_t;\mathcal{D})$

  \If{$t \bmod K = 0$}
    \State Estimate local advantages $A^\pi_i(x,a)$ using $Q_{\theta^Q_{t+1}}$
    \State Estimate reach probabilities $\rho(x)$ using $\rho_{\theta^\rho_{t+1}}$

    \State Compute logit update direction:
    \Statex \quad\quad\quad\quad $\Delta L(a|x) = \rho_{\theta^\rho_{t+1}}(x)\,
      \tfrac{A^\pi_i(x,a)}{\pi_{\theta^L_t}(a|x)}$

    \State Update actor: 
    $\theta^L_{t+1} \gets \theta^L_t + \eta_t \nabla_{\theta^L} \langle L_{\theta^L_t}(x), \Delta L(x)\rangle$

    \State Update policy: $\pi_{t+1}(a|x) = \mathrm{softmax}(L_{\theta^L_{t+1}}(x))$
  \EndIf
\EndFor
\end{algorithmic}
\end{algorithm}

\section{Experiments}\label{sec:exp}
For validation, we benchmark our proposed BNN-based framework against state-of-the-art regularized RD-based approaches in both normal-form and extensive-form games. In each game, we add unbiased noise directly into the rewards in the normal-form case or the counterfactual values in the extensive-form case. To this end, we consider Biased Rock–Paper–Scissors (BRPS) and its four-action variant (BRPS-W), as well as Kuhn Poker and Leduc Poker. To further test adaptability, we construct non-stationary versions of BRPS and Kuhn Poker, where the underlying payoffs change either instantaneously (direct-shift mode) or gradually over time (continuous-shift mode).
As a baseline, we compare against regularized RD-based approaches, including adaptively perturbed mirror descent (APMD)~\cite{abe2023adaptively} for normal-form games and R-Nad~\cite{perolat2021poincare} for extensive-form games. They are both state-of-the-art algorithms that achieve last-iterate convergence in those settings.

In stationary games, our method exhibits markedly faster convergence and smoother trajectories throughout training, achieving low exploitability earlier and maintaining stability over extended horizons. Whereas the baseline occasionally attains a lower asymptotic \textsc{NashConv}, this difference corresponds to the noise plateau predicted by our theoretical analysis. Under non-stationary conditions, our approach demonstrates superior adaptability: it rapidly converges after abrupt payoff changes and sustains low exploitability during gradual transitions.

\begin{figure}[b]
    \centering
    \includegraphics[width=\linewidth]{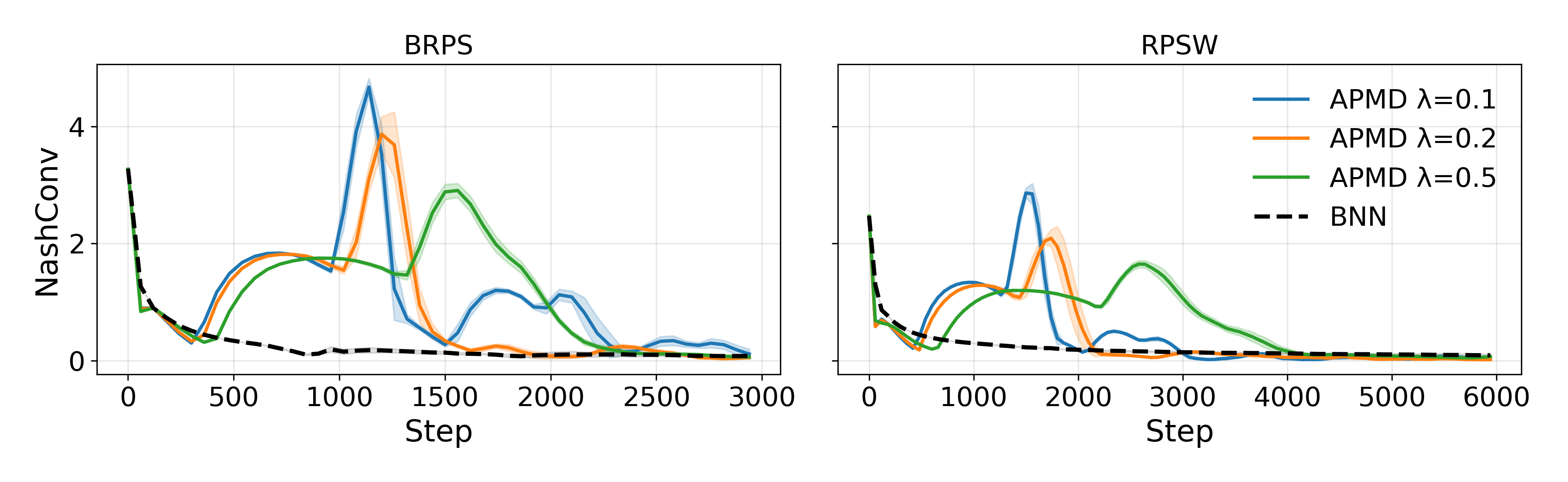}
    \caption{BRPS and BRPS-W.}
    \label{fig:BRPS}
\end{figure}

\begin{figure}[t]
    \centering
    \includegraphics[width=\linewidth]{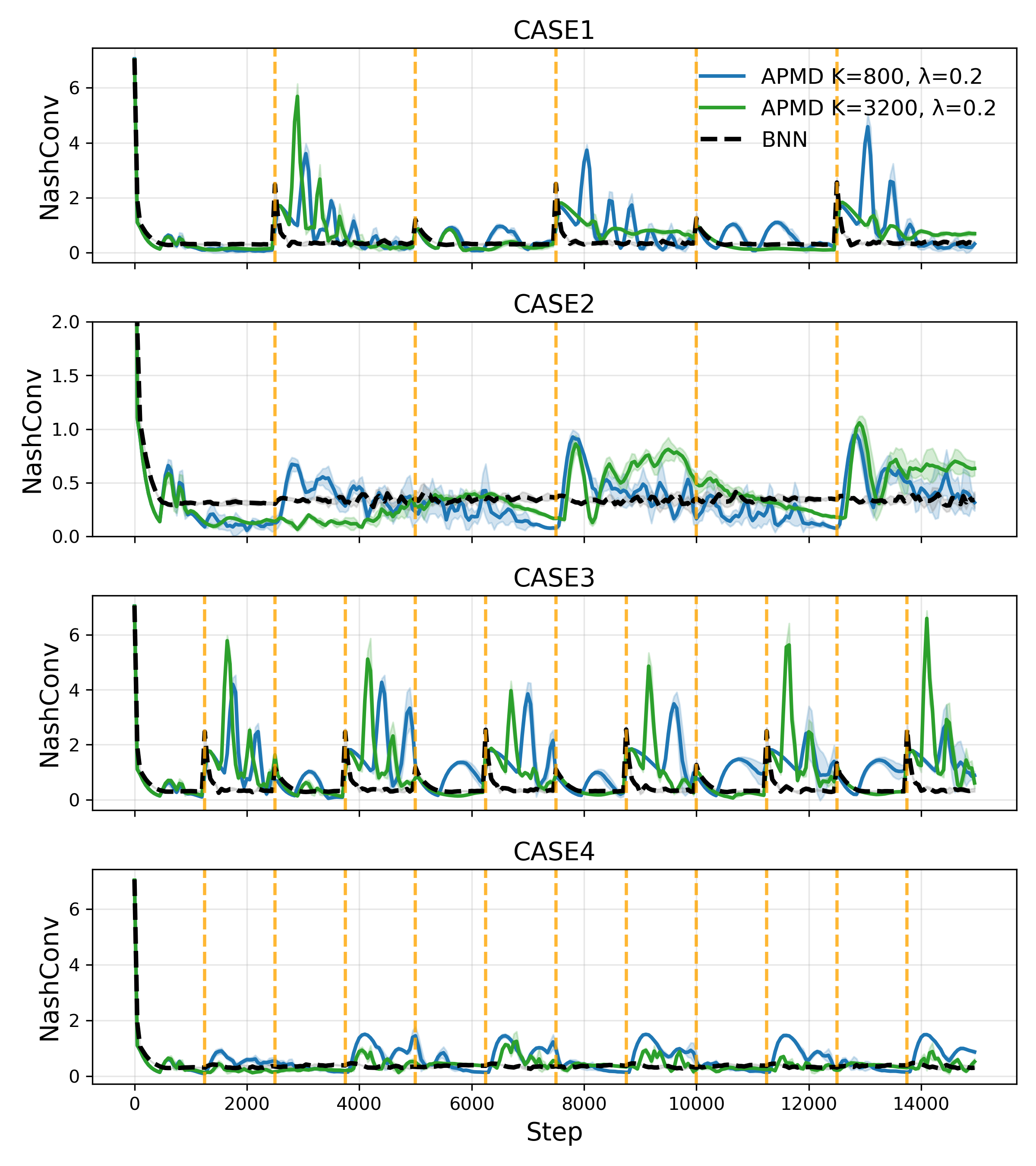}
    \caption{Nonstationary RPS.}
    \label{fig:nonsRPS}
\end{figure}

\textbf{Normal-form Games}. The RPS family of games is defined by three parameters controlling the relative advantages between the pairwise matchups $(a_{RP}, a_{PS}, a_{SR})$. In the BRPS environment, $a_{RP}$ is fixed at $12$, creating a persistent bias that yields a unique interior Nash equilibrium. The BRPS-W variant extends this setting with a fourth action that interacts neutrally with all existing ones, producing payoffs of zero against every opponent action. 
For these experiments, we revert to a purely tabular version of the algorithm without any function approximation. This design isolates the effect of the underlying dynamics from the influence of neural approximation, allowing a direct comparison between the methods. 

As depicted in Fig.~\ref{fig:BRPS}, the BNN-based method maintains low exploitability and stable convergence throughout training until it reaches the noise plateau. Once the trajectory stabilizes near this region, the confidence intervals widen slightly in the later stages, indicating a modest increase in variance. This effect is consistent with our theoretical analysis, which predicts that the influence of stochastic noise becomes amplified only in the close vicinity of the Nash equilibrium. In contrast, the APMD displays large confidence intervals in early stages of training. These occur when its trajectories move too close to the simplex boundary, causing unstable probability mass distributions. This edge concentration leads to substantial fluctuations and irregular convergence behavior.

To assess adaptability under changing environments, we test a non-stationary extension of the Biased Rock–Paper–Scissors game in which the payoff parameters vary over time while preserving the zero-sum game structure shown in Fig.~\ref{fig:nonsRPS}. The three matchup coefficients $(a_{RP}, a_{PS}, a_{SR})$ change sequentially such that the bias rotates between matchups:
$(12, 1, 1) \rightarrow (6.5, 6.5, 1) \rightarrow (1, 12, 1)$. 
We consider four variants of this process. In case 1, each stage lasts for $2500$ steps. 
In case 2, the game first remains fixed for $2500$ steps, and the subsequent transition between configurations is made gradually over another $2500$ steps. 
Cases 3 and 4 follow the same patterns as cases 1 and 2, respectively, but each stage lasts for only $1250$ steps. 
In this setting, we further evaluate the APMD under different hyperparameter configurations, varying the regularization update frequency ($K$) and the regularization strength ($\lambda$). Our BNN-based algorithm maintains low \textsc{NashConv} after each environmental transition, showing stable adaptation across all tested cases. In comparison, APMD adapts more slowly and exhibits transient instability following each change. Its performance also differs noticeably across transition speeds. 
Moreover, different combinations of hyperparameters yield inconsistent outcomes across transition modes, indicating that no single configuration performs robustly. 

\begin{figure}[!t]
\begin{minipage}[h]{1.0\linewidth}
\centering
\includegraphics[width=\linewidth]{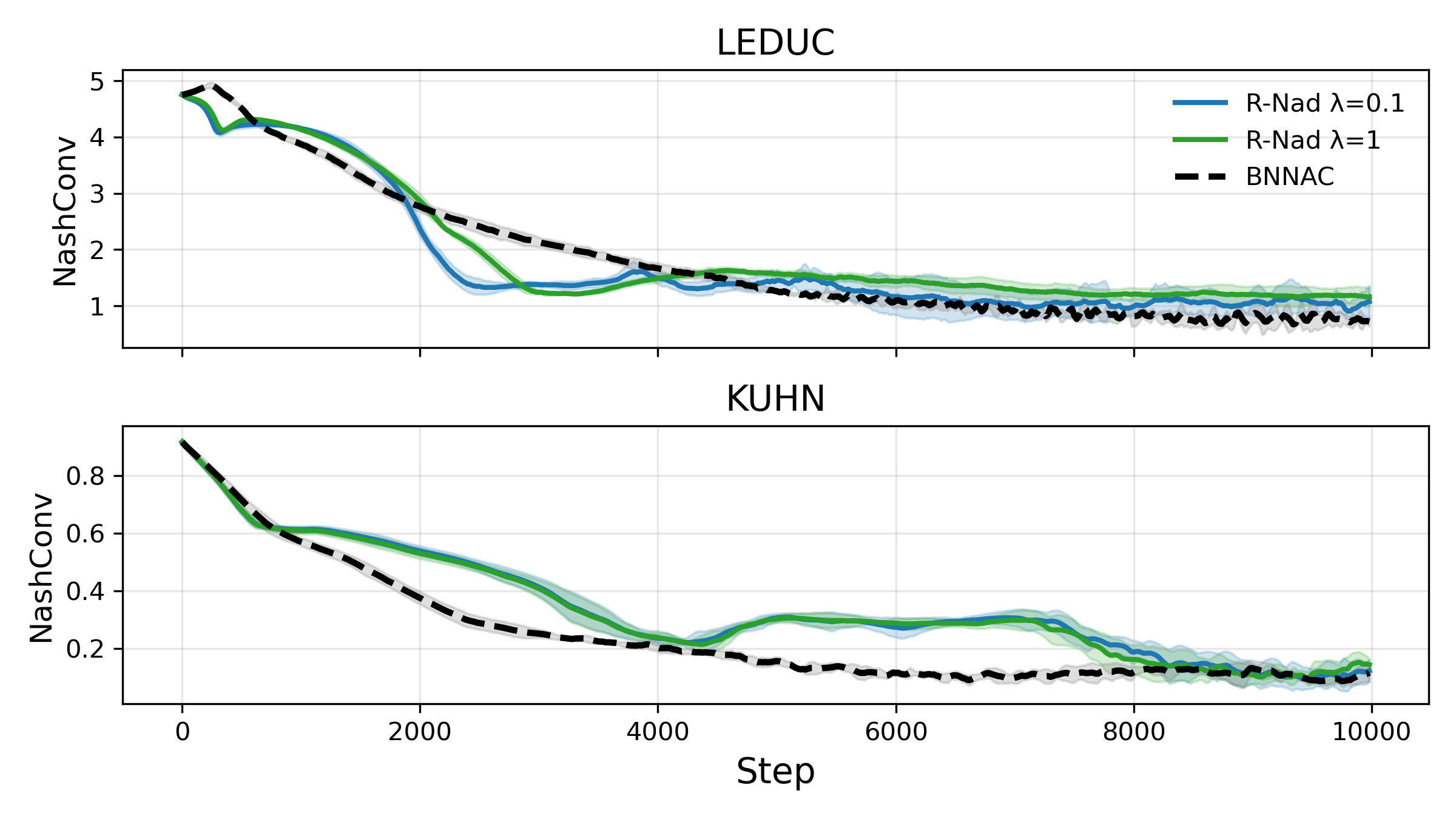}
\end{minipage}
\vspace{0.00mm} 
\begin{minipage}[h]{1.0\linewidth}
\centering
\includegraphics[width=\linewidth]{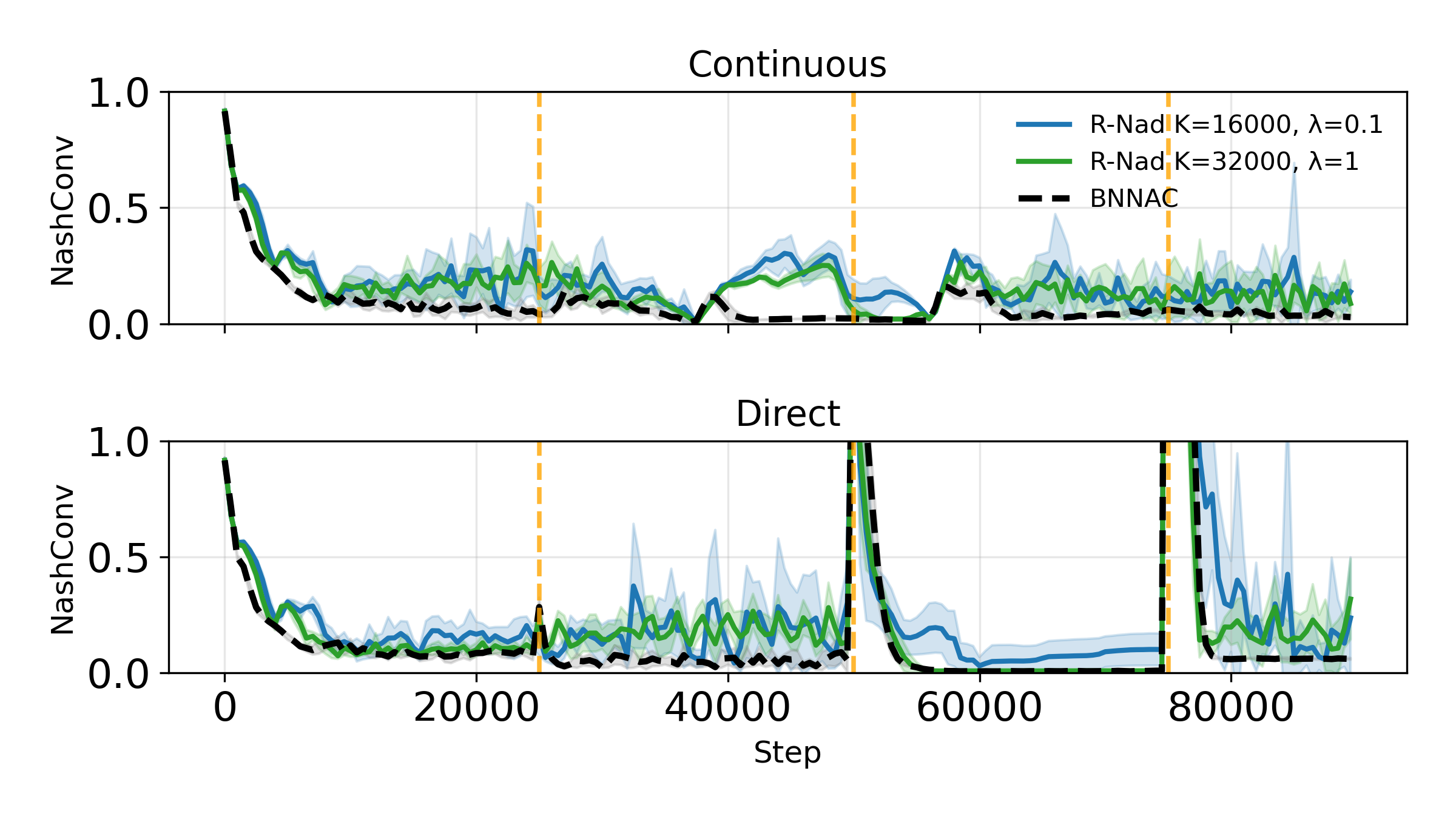}
\end{minipage}   


\caption{Stationary and nonstationary poker games.}
\label{fig:poker}
\end{figure}


\textbf{Extensive-form Games}. 
We now consider stationary Kuhn and Leduc poker, and also construct two non-stationary variants of Kuhn Poker, in which the underlying payoffs evolve gradually over time as shown in Fig.~\ref{fig:poker}. 
For all experiments, we employ the full neural implementation of our framework, called BNNAC, where both the actor and critic networks are trained with reach-weighted returns. The state-of-the-art baseline, R-NaD, is implemented under the same network architecture and optimization settings in the noisy environment. For all the environments, we sweep the regularization strength to examine the sensitivity of R-NaD to the rate and magnitude of environmental change. In the stationary settings, both algorithms exhibit rapid early-stage improvement in \textsc{NashConv}, with BNNAC showing a slightly faster initial descent. Overall, the results confirm that both approaches demonstrate stable convergence in extensive-form games under stationary conditions. 

In the non-stationary settings, we modify the bet size over three distinct stages, changing from the initial value of $1$ to $2$, then to $-2$, and finally to $6$. In the continuous variant, the bet size smoothly returns to $1$ in the last stage. Across all cases, we observe that the current strategies experience inevitable deviations in \textsc{NashConv} whenever the environment undergoes a major transition. Under these conditions, the two algorithms exhibit qualitatively different behaviors. While neither method can completely offset the environmental shifts, BNNAC consistently achieves lower \textsc{NashConv}, whereas R-NaD displays unstable oscillatory patterns. 

\section{Conclusion}\label{sec:conclusion}
This work revisits the classical BNN dynamics and establishes its theoretical and practical relevance in modern multi-agent learning. We show that BNN dynamics guarantee regularization-free last-iterate convergence in both normal-form and extensive-form zero-sum games, maintaining stability under noisy feedback. The resulting BNNAC algorithm scales to neural implementations and consistently outperforms baselines in both stationary and non-stationary environments. Future work will focus on accelerating convergence and further mitigating the effect of stochastic noise to enhance performance in large-scale and highly dynamic settings.



\begin{acks}
\texttt{LS and TZ have been supported by the Advanced Research and Invention Agency (ARIA), Safeguarded AI programme.}
\end{acks}



\bibliographystyle{ACM-Reference-Format} 
\bibliography{sample}

\clearpage
\onecolumn

\input{appendix}

\end{document}

%% file: Appendix.tex
\def\tz#1{{\color{magenta}#1}}

\section*{Appendix}
\addcontentsline{toc}{section}{Appendix}
\section{Theoretical proof}
\subsection{Proof of Lemma 1}

Define the per-action Jensen gap
\[
\delta(a):=\mathbb{E}\bigl[[\mathrm{Adv}(a)+\varepsilon(a)]_+\bigr]-[\mathrm{Adv}(a)]_+ .
\]
Since $[\cdot]_+$ is convex, $\delta(a)\ge 0$. Moreover, $[\cdot]_+$ is $1$-Lipschitz, hence
for any $x$ and any real $\eta$,
\[
\bigl|[x+\eta]_+-[x]_+\bigr|\le |\eta|.
\]
Taking the expectation yields
\[
0\le \delta(a)\le \mathbb{E}|\varepsilon(a)|
\le \sqrt{\mathbb{E}[\varepsilon(a)^2]}\le \sigma,
\]
where we used Cauchy--Schwarz in the second inequality.

Now recall that the bias term $\beta(\pi)$ is defined by
$\mathbb{E}[H(\pi)]=H(\pi)+\beta(\pi)$, and its coordinate form is
\[
\beta(a)=\delta(a)-\pi(a)\sum_{b\in A_i}\delta(b).
\]
Therefore, we obtain
\[
|\beta(a)|
\le (1-\pi(a))\sigma+\pi(a)(|A_i|-1)\sigma
= \sigma\bigl(1+\pi(a)(|A_i|-2)\bigr)
\le (|A_i|-1)\sigma.
\]
\subsection{Proof of Lemma 2}

Recall the stochastic approximation decomposition
\[
\pi_{t+1}=\pi_t+\eta_t\bigl(H(\pi_t)+\beta(\pi_t)+\zeta_{t+1}\bigr),
\]
where $\mathbb{E}[\zeta_{t+1}\mid \pi_t]=0$ and $\|\beta(\pi)\|_\infty\le (|A_i|-1)\sigma$
by Lemma~1.

Let $r(\pi):=\Delta_+(\pi)$ denote the vector with coordinates
$r_a(\pi)=[u_i(a;\pi)-u_i(\pi)]_+$, and recall
\[
\Gamma(\pi)=\frac12\|r(\pi)\|_2^2,
\qquad
S(\pi)=\sum_{a\in A_i} r_a(\pi)=\|r(\pi)\|_1.
\]
For the noiseless flow $\dot \pi=H(\pi)$ we have the dissipation identity
\[
\frac{d}{dt}\Gamma(\pi)=-2S(\pi)\Gamma(\pi),
\]
and the elementary inequality $S(\pi)^2=\|r(\pi)\|_1^2\ge \|r(\pi)\|_2^2=2\Gamma(\pi)$.

We now derive a one-step bound for the discrete recursion.
By a second-order Taylor expansion of $\Gamma$ along the increment
$\eta_t\bigl(H(\pi_t)+\beta(\pi_t)+\zeta_{t+1}\bigr)$, there exists a finite
constant $C_3>0$ (depending only on the game) such that
\begin{align*}
\Gamma(\pi_{t+1})
&\le \Gamma(\pi_t)
+\eta_t\Bigl(-2S(\pi_t)\Gamma(\pi_t)\Bigr)
+\eta_t\bigl\langle r(\pi_t),\,\beta(\pi_t)+\zeta_{t+1}\bigr\rangle
+ C_3\eta_t^2 .
\end{align*}
Taking conditional expectation given $\pi_t$ and using
$\mathbb{E}[\zeta_{t+1}\mid \pi_t]=0$ yields
\begin{align*}
\mathbb{E}\bigl[\Gamma(\pi_{t+1})\mid \pi_t\bigr]
\le \Gamma(\pi_t)
-2\eta_t S(\pi_t)\Gamma(\pi_t)
+\eta_t\bigl\langle r(\pi_t),\,\beta(\pi_t)\bigr\rangle
+ C_3\eta_t^2 .
\end{align*}
Next, we bound the bias inner product by Hölder:
\[
\bigl|\langle r(\pi_t),\beta(\pi_t)\rangle\bigr|
\le \|r(\pi_t)\|_1\|\beta(\pi_t)\|_\infty
= S(\pi_t)\|\beta(\pi_t)\|_\infty
\le S(\pi_t)(|A_i|-1)\sigma .
\]
Therefore,
\[
\mathbb{E}\bigl[\Gamma(\pi_{t+1})\mid \pi_t\bigr]
\le \Gamma(\pi_t)
-2\eta_t S(\pi_t)\Gamma(\pi_t)
+(|A_i|-1)\sigma\,\eta_t\,S(\pi_t)
+ C_3\eta_t^2 .
\]
Finally, using $S(\pi_t)\ge \sqrt{2\Gamma(\pi_t)}$ and
$S(\pi_t)=\|r(\pi_t)\|_1\le \sqrt{|A_i|}\|r(\pi_t)\|_2=\sqrt{2|A_i|\Gamma(\pi_t)}$,
we obtain
\[
\mathbb{E}\bigl[\Gamma(\pi_{t+1})\mid \pi_t\bigr]
\le \Gamma(\pi_t)
-2\sqrt{2}\,\eta_t\,\Gamma(\pi_t)^{3/2}
+(|A_i|-1)\sqrt{2|A_i|}\,\sigma\,\eta_t\,\Gamma(\pi_t)^{1/2}
+ C_3\eta_t^2 .
\]
Taking full expectations and denoting $g_t:=\mathbb{E}[\Gamma(\pi_t)]$
gives the claimed inequality.

\subsection{Proof of Theorem 1}

From Lemma~2, the expected one–step drift of the Lyapunov potential
satisfies
\begin{align*}
  g_{t+1}
  &\le g_t
   - 2\sqrt{2}\,\eta_t\, g_t^{3/2}
   + (|A_i|-1)\sqrt{2|A_i|}\,\sigma\,\eta_t\, g_t^{1/2}
   + C_3\eta_t^2,
\end{align*}
where $g_t = \mathbb{E}[\Gamma(\pi_t)]$.
The last term $C_3\eta_t^2$ is of higher order in the step size.
Because $\sum_t \eta_t^2 < \infty$ under the Robbins--Monro
conditions, its cumulative effect is finite and does not affect the
asymptotic behaviour, so it can be neglected in determining the steady
state of the mean drift.

Setting the remaining leading terms to equilibrium gives
\[
  - 2\sqrt{2}\, g^{3/2}
  + (|A_i|-1)\sqrt{2|A_i|}\,\sigma\, g^{1/2} = 0 ,
\]
whose positive root is
\[
  g_* = \frac{(|A_i|-1)\sqrt{|A_i|}}{2}\,\sigma .
\]
The expected drift is negative whenever $g_t > g_*$ , implying that $\{g_t\}$ converges almost surely to an
$O(\sigma)$ neighbourhood of this equilibrium.
Applying the Robbins--Siegmund supermartingale theorem under
$\sum_t\eta_t=\infty$ and $\sum_t\eta_t^2<\infty$ yields
\[
  \limsup_{t\to\infty} \Gamma(\pi_t)
  \le g_* = \tfrac12 (|A_i|-1)\sqrt{|A_i|}\,\sigma
  \quad\text{almost surely.}
\]
Therefore, the stochastic BNN recursion stabilises within an
$O(\sigma)$ neighbourhood of the Nash set, completing the proof.

\subsection{Proof of Theorem~2}

From Lemma~2, the expected Lyapunov potential $g_t=\mathbb{E}[\Gamma(\pi_t)]$
satisfies, up to higher-order $O(\eta_t^2)$ terms,
\[
  g_{t+1} \le g_t - a\,\eta_t\,g_t^{3/2} + c\,\eta_t^2,
  \qquad  a,c>0.
\]
The $O(\eta_t^2)$ term is summable under $\sum_t\eta_t^2<\infty$
and therefore does not affect the asymptotic rate.
Introducing $y_t = 1/\sqrt{g_t}$ so that $g_t = y_t^{-2}$,
we have
\[
  y_{t+1}^{-2} - y_t^{-2} = g_{t+1} - g_t
  \le -a\,\eta_t\,y_t^{-3} + c\,\eta_t^2 .
\]
Expanding the left-hand side via
$y_{t+1}^{-2}-y_t^{-2} = -2y_t^{-3}(y_{t+1}-y_t) + O((y_{t+1}-y_t)^2)$
and neglecting the $O(\eta_t^2)$ remainder yields
\[
  y_{t+1}-y_t \ge \tfrac{a}{2}\eta_t - K\eta_t^2
\]
for some finite constant $K>0$. Summing from an initial index $t_0$
to $t-1$ gives
\[
  y_t - y_{t_0}
  \ge \tfrac{a}{2}\sum_{k=t_0}^{t-1}\eta_k
    - K\sum_{k=t_0}^{t-1}\eta_k^2 .
\]
The second sum converges, while the first dominates the growth of
$y_t$. With $\eta_t = c(t+t_0)^{-2/3}$, the integral comparison
$\sum_{k\le t} k^{-2/3} = \Theta(t^{1/3})$ implies that
$y_t = \Theta(t^{1/3})$ as $t\to\infty$.
Since $g_t = y_t^{-2}$, we obtain
\[
  \mathbb{E}[\Gamma(\pi_t)] = g_t = O(t^{-2/3})
\]
in the noiseless case.
When $\sigma>0$, the additional bias term
$b\sigma\eta_t g_t^{1/2}$ in Lemma~2 modifies the drift so that the
decay halts once $a g_t^{3/2}\!\approx\! b\sigma g_t^{1/2}$,
producing an $O(\sigma)$ steady-state floor.
Combining the transient and stationary regimes yields
\[
  \mathbb{E}[\Gamma(\pi_t)] = O(t^{-2/3}) + O(\sigma),
\]
which establishes the claimed convergence rate.

\subsection{Proof of Theorem~3}

At a fixed point of the biased BNN dynamics, the stationary condition $\dot{\pi}(a) = 0$ implies
\[
  [u(a)-\bar{u}]_+ + \delta(a)
  = \pi(a)\sum_{b\in A}\bigl([u(b)-\bar{u}]_+ + \delta(b)\bigr),
  \quad \forall a\in\pi,
\]
or equivalently,
\[
  \frac{[u(a)-\bar{u}]_+ + \delta(a)}{\pi(a)} = \kappa,
  \quad \forall a\in(\pi),
\]
where $\kappa>0$ is a normalising constant. The bias term $\delta(a)$ is defined by Gaussian smoothing:
\[
  \delta(a) := \mathbb{E}\bigl[[u(a)-\bar{u}+\varepsilon]_+\bigr] - [u(a)-\bar{u}]_+, \qquad \varepsilon\sim\mathcal{N}(0,\sigma^2).
\]
Because of this definition, we have $[u(a)-\bar{u}]_+ + \delta(a) > 0$ for all $a$, and the mapping
$[u(a)-\bar{u}]_+ \mapsto [u(a)-\bar{u}]_+ + \delta(a)$ is strictly increasing.

Now consider two actions $a_1,a_2\in(\pi)$ with $\pi(a_1)>\pi(a_2)$. From the fixed-point equation,
\[
  \frac{[u(a_1)-\bar{u}]_+ + \delta(a_1)}{\pi(a_1)}
  = \frac{[u(a_2)-\bar{u}]_+ + \delta(a_2)}{\pi(a_2)}.
\]
Since $\pi(a_1)>\pi(a_2)$, this implies
\[
  [u(a_1)-\bar{u}]_+ + \delta(a_1) > [u(a_2)-\bar{u}]_+ + \delta(a_2).
\]
Because the mapping $[u(a)-\bar{u}]_+\mapsto [u(a)-\bar{u}]_+ + \delta(a)$ is strictly increasing, we obtain
\[
  [u(a_1)-\bar{u}]_+ > [u(a_2)-\bar{u}]_+.
\]
Hence the advantage values $[u(a)-\bar{u}]_+$ are strictly decreasing in the ranking of $\pi(a)$,
that is, actions with smaller probability mass must have lower advantage.

In the degenerate case where $\pi(a)$ is uniform over its support, the fixed-point equation implies
\[
  [u(a)-\bar{u}]_+ + \delta(a) = \text{const}, \quad \forall a\in(\pi).
\]
Since $\delta(a)$ is strictly increasing in $[u(a)-\bar{u}]_+$, this is only possible if all
$[u(a)-\bar{u}]_+$ are equal. Together with the normalization
$\sum_a \pi(a)[u(a)-\bar{u}]_+ = 0$, this implies
\[
  u(a) - \bar{u} = 0, \quad \forall a\in A,
\]
i.e., $\pi$ corresponds exactly to a Nash equilibrium.

In the non-degenerate case, $\pi$ is non-uniform, so the advantages are not all equal. Since
$[u(a)-\bar{u}]_+$ is strictly decreasing in $\pi(a)$, we have:
\begin{itemize}
  \item at least one action with strictly positive advantage,
  \item at least one action with strictly negative advantage.
\end{itemize}
Let $a_{\max}$ denote the action with the largest probability mass (and thus largest advantage):
\[
  a_{\max} := \arg\max_a \pi(a), \quad
  u_{\max} := u(a_{\max}), \quad
  \pi_{\max} := \pi(a_{\max}).
\]
Define the subset of negatively advantaged actions
\[
  A^- := \{a\in(\pi)\mid u(a)-\bar{u}<0\}, \qquad m := |A^-|,
\]
the total probability mass
\[
  \pi^- := \sum_{a\in A^-}\pi(a), \qquad
  \pi^-_{\max} := \max_{a\in A^-}\pi(a),
\]
and the average negative advantage
\[
  \bar{u}^- := \frac{1}{\pi^-}\sum_{a\in A^-}\pi(a)(u(a)-\bar{u}).
\]
By the zero-sum identity $\sum_a \pi(a)(u(a)-\bar{u})=0$, we have
\[
  (u_{\max}-\bar{u})\pi_{\max} \le -\pi^-\,\bar{u}^-.
\]
Bounding $\pi^- \le m\,\pi^-_{\max}$ yields
\[
  u_{\max}-\bar{u} \le m\,\frac{\pi^-_{\max}}{\pi_{\max}}(-\bar{u}^-).
\]
From the positivity of $[u(a)-\bar{u}]_+ + \delta(a)$ and the definition of $\delta(a)$,
we have $u(a)-\bar{u} > -\delta(a)$, so $-\bar{u}^- < \delta_{\max}$.
Using $\delta_{\max} = \sigma/\sqrt{2\pi}$ from Lemma~2 and
$\pi^-_{\max}\le \pi_{\max}$ gives
\[
  u_{\max}-\bar{u} < m\frac{\sigma}{\sqrt{2\pi}}.
\]

Substituting this bound into the Lyapunov potential
$\Gamma(\pi) = \tfrac12\sum_a [u(a)-\bar{u}]_+^2$ yields
\[
  \Gamma(\pi')
  < \tfrac12 |A_i| \left(m\,\frac{\sigma}{\sqrt{2\pi}}\right)^2
  < \tfrac12 |A_i|(|A_i|-1)^2 \frac{\sigma^2}{2\pi}
  = O(\sigma^2).
\]
This shows that the steady-state Lyapunov potential under biased BNN dynamics
scales quadratically with the perturbation magnitude~$\sigma$,
completing the proof.

\subsection{Proof of Lemma 3}

Fix a player $i$ and an information set $x \in \mathcal I_i$.
Define the counterfactual value vector
$$
F_i^x(\pi)
:= \rho_{-i}^\pi(x)\, v_i^\pi(x,\cdot),
$$
and the corresponding counterfactual advantage vector
$$
\hat F_i^x(\pi)
:= F_i^x(\pi)
- \big(\pi_i(\cdot \mid x)^\top F_i^x(\pi)\big)\mathbf 1
= \rho_{-i}^\pi(x)\, A_i^\pi(x,\cdot).
$$
Define the information--set Lyapunov function
$$
\Gamma_i^x(\pi)
:= \frac12 \big\| \hat F_{i,+}^x(\pi) \big\|^2
= \frac12 \big\| \rho_{-i}^\pi(x)\, A_{i,+}^\pi(x) \big\|^2,
$$
and
$$
S_i^x(\pi)
:= \mathbf 1^\top \rho_{-i}^\pi(x)\, A_{i,+}^\pi(x).
$$

Along a trajectory $\pi(t)$, the time derivative of $\Gamma_i^x(\pi)$ satisfies
$$
\begin{aligned}
\dot{\Gamma}_i^x(\pi)
&= \nabla_\pi \Gamma_i^x(\pi)^\top \dot\pi
\\
&= \hat F_{i,+}^x(\pi)^\top \, D\hat F_i^x(\pi)\, \dot\pi
\\
&= \hat F_{i,+}^x(\pi)^\top
\Big(
DF_i^x(\pi)
- \mathbf 1 \big(
\pi_i(\cdot \mid x)^\top DF_i^x(\pi)
+ F_i^x(\pi)^\top
\big)
\Big)\dot\pi
\\
&=
\dot\pi_i(\cdot \mid x)^\top \, DF_i^x(\pi)\, \dot\pi
- \big(\mathbf 1^\top \hat F_{i,+}^x(\pi)\big)
\big(F_i^x(\pi)^\top \dot\pi\big)
\\
&=
\dot\pi_i(\cdot \mid x)^\top \, DF_i^x(\pi)\, \dot\pi
- S_i^x(\pi)\cdot 2\Gamma_i^x(\pi).
\end{aligned}
$$

Then the global Lyapunov function as the sum of information--set Lyapunov functions over all players and information sets,
$$
V(\pi)
:= \sum_{i}\; \sum_{x \in \mathcal I_i} \Gamma_i^x(\pi).
$$
Along a trajectory $\pi(t)$, its time derivative satisfies
$$
\begin{aligned}
\dot V(\pi)
&= \sum_{i}\; \sum_{x \in \mathcal I_i} \dot{\Gamma}_i^x(\pi)
\\
&= \sum_{i}\; \sum_{x \in \mathcal I_i}
\Big(
\dot\pi_i(\cdot \mid x)^\top \, DF_i^x(\pi)\, \dot\pi
- S_i^x(\pi)\cdot 2\Gamma_i^x(\pi)
\Big).
\end{aligned}
$$

Under the zero--sum condition, the counterfactual value fields satisfy
$$
\sum_i \sum_{x \in \mathcal I_i}
\dot\pi_i(\cdot \mid x)^\top \, DF_i^x(\pi)\, \dot\pi
= 0.
$$
Consequently, the time derivative of the global Lyapunov function reduces to
$$
\begin{aligned}
\dot V(\pi)
&=
\sum_{i}\; \sum_{x \in \mathcal I_i}
\Big(
\dot\pi_i(\cdot \mid x)^\top \, DF_i^x(\pi)\, \dot\pi
- S_i^x(\pi)\cdot 2\Gamma_i^x(\pi)
\Big)
\\
&=
- \sum_{i}\; \sum_{x \in \mathcal I_i}
S_i^x(\pi)\cdot 2\Gamma_i^x(\pi).
\end{aligned}
$$

\subsection{Proof of Lemma~4 and Lemma~5}

The proofs follow the same decomposition and Lyapunov argument as in the
normal-form case.
The only difference is that payoffs are replaced by local counterfactual
advantages weighted by opponent reach probabilities, so the resulting constants
depend on the information-set structure.

\paragraph{Lemma~4.}
As in Lemma~1, the convexity and $1$-Lipschitz continuity of $[\cdot]_+$ imply a
uniform bound on the structural bias.
Moreover, since the payoff noise is unbiased with bounded variance
(Assumption~1), the induced martingale-difference noise has uniformly bounded
second moment.
Therefore, there exist constants $C_1,C_2>0$, depending only on the local action
set sizes, such that
\[
\|\beta'(\pi)\|\le C_1\sigma,
\qquad
\mathbb E[\|\zeta'_{k+1}\|^2\mid \pi_k]\le C_2\sigma^2.
\]

\paragraph{Lemma~5.}
Using the stochastic approximation representation
\[
\pi_{k+1}
=\pi_k+\alpha_k\bigl(H'(\pi_k)+\beta'(\pi_k)+\zeta'_{k+1}\bigr),
\]
a second-order expansion of the Lyapunov potential $V$ and Lemma~3 yield the
deterministic descent term $-2\rho\,\alpha_k V(\pi_k)^{3/2}$.
The bias and noise contributions are controlled by Lemma~4, while the quadratic
remainder contributes an $O(\alpha_k^2)$ term.
Taking expectations gives
\[
g_{k+1}
\le g_k
-2\rho\,\alpha_k g_k^{3/2}
+C_1\sigma\,\alpha_k g_k^{1/2}
+C_2\alpha_k^2,
\]
where $g_k=\mathbb E[V(\pi_k)]$, which proves Lemma~5.

\subsection{Proof of Theorem~4 and Theorem~5}

The proofs of Theorems~4 and~5 follow the same analytical structure as
Theorems~2 and~3. 
While the arguments for Theorems~2 and~3 could have been presented jointly,
they were written separately for clarity since they first introduced the
core Lyapunov analysis. 
As Theorems~4 and~5 employ the same reasoning in the value-based
extensive-form setting, we present their proofs together to avoid unnecessary
repetition.

From Lemma~5, the expected one-step change of the value-based Lyapunov
potential satisfies
\[
  g_{k+1} - g_k
  \le -a\,\alpha_k\,g_k^{3/2}
      + c_1\sigma\,\alpha_k\,g_k^{1/2}
      + c_2\alpha_k^2 ,
\]
where $g_k=\mathbb{E}[V(\pi_k)]$, and the constants
$a,c_1,c_2>0$ depend only on the reach-probability bounds and the
smoothness of the value function.
The first term drives the monotone decrease of $V(\pi)$, the second term
captures the stochastic bias, and the third is a higher-order step-size term.

Because $g_k=O(1)$ during the transient phase,
the bias term $c_1\sigma\alpha_k g_k^{1/2}$ is asymptotically dominated by
the descent term $a\alpha_k g_k^{3/2}$.
Hence the stochastic perturbation does not affect the convergence rate,
and the same first-order expansion used in Theorem~2 applies.
Let $y_k=g_k^{-1/2}$.
Then, up to $O(\alpha_k^2)$ terms,
\[
  y_{k+1}-y_k \ge \tfrac{a}{2}\alpha_k - K\alpha_k^2
\]
for some constant $K>0$.
Summing over $k$ with $\alpha_k=O(k^{-2/3})$ yields
\[
  y_k = \Theta(k^{1/3}) + O(1),
  \qquad
  g_k = O(k^{-2/3}) .
\]

As $g_k$ decreases, the descent and bias terms eventually balance at
$a g_k^{3/2}\approx c_1\sigma g_k^{1/2}$,
leading to a steady-state level $g_k = O(\sigma)$.
Combining the transient decay and the stationary bias gives the unified
bound
\[
  \mathbb{E}[V(\pi_k)]
   = O(k^{-2/3}) + O(\sigma) .
\]
Therefore, the value-based Lyapunov potential in the extensive-form
game decays with the same asymptotic rate as in the normal-form case,
and the stochastic perturbation only determines the steady-state noise floor
without changing the convergence rate.
This completes the proofs of Theorems~4 and~5.

\subsection{Proof of Theorem~6}

The argument mirrors that of Theorem~3 but for the value-based potential
in the extensive-form setting. 
At the stationary point of the biased BNN dynamics, the expected update
vanishes at every information set $h\in\mathcal{H}_i$:
\[
  \frac{[u_i(h,a)-\bar u_i(h)]_+ + \delta_i(h,a)}{\pi_i(a|h)} = \kappa_i(h),
  \qquad \forall a\in A(h),
\]
where $\kappa_i(h)>0$ is a normalising constant and
$\delta_i(h,a)$ is the stochastic perturbation caused by payoff noise.
By Lemma~5, the perturbation satisfies
$|\delta_i(h,a)|\le \delta_{\max}=O(\sigma)$
and is strictly increasing in $u_i(h,a)$.

\paragraph{Step~1. Local monotonicity.}
For any two actions $a_1,a_2\in A(h)$,
\[
  \frac{[u_i(h,a_1)-\bar u_i(h)]_+ + \delta_i(h,a_1)}{\pi_i(a_1|h)}
  = \frac{[u_i(h,a_2)-\bar u_i(h)]_+ + \delta_i(h,a_2)}{\pi_i(a_2|h)} .
\]
If $\pi_i(a_1|h)>\pi_i(a_2|h)$, then by the monotonicity of $\delta_i$
it follows that
$[u_i(h,a_1)-\bar u_i(h)]_+>[u_i(h,a_2)-\bar u_i(h)]_+$,
i.e., the stationary policy preserves the ordering of local advantages.

\paragraph{Step~2. Bounding local advantages.}
Within each $h$, let $a_{\max}(h)$ denote the action with the largest
probability and hence largest positive advantage
$u_{i,\max}(h)-\bar u_i(h)$.
Let $A^-(h)=\{a:u_i(h,a)-\bar u_i(h)<0\}$ and $m(h)=|A^-(h)|$.
From the zero-mean property
$\sum_a \pi_i(a|h)[u_i(h,a)-\bar u_i(h)]=0$,
we obtain
\[
  (u_{i,\max}(h)-\bar u_i(h))\pi_{i,\max}(h)
   \le -\pi^-_i(h)\,\bar u_i^-(h),
\]
and hence
\[
  u_{i,\max}(h)-\bar u_i(h)
   \le m(h)\frac{\pi^-_{i,\max}(h)}{\pi_{i,\max}(h)}(-\bar u_i^-(h)).
\]
Using $-\bar u_i^-(h)<\delta_{\max}=O(\sigma)$
and $\pi^-_{i,\max}(h)\le\pi_{i,\max}(h)$ gives
\[
  u_{i,\max}(h)-\bar u_i(h)
   < m(h)\frac{\sigma}{\sqrt{2\pi}} .
\]

\paragraph{Step~3. Aggregating over information sets.}
The value-based Lyapunov potential sums the squared positive advantages
over all information sets, weighted by the player’s reach probabilities:
\[
  V_i(\pi_i)
  = \sum_{h\in\mathcal{H}_i}\rho_i(h)
    \sum_{a\in A(h)} [u_i(h,a)-\bar u_i(h)]_+^2 .
\]
Applying the bound above to each $h$ yields
\[
  V_i(\pi_i')
   \le \frac12
      \sum_{h\in\mathcal{H}_i}
        \rho_i(h)\,|A(h)|\bigl(m(h)\,\tfrac{\sigma}{\sqrt{2\pi}}\bigr)^2
   = O(\sigma^2).
\]
Summing over all players gives
\[
  \mathbb{E}[V(\pi')] = O(\sigma^2),
\]
which characterises the deviation of the fixed state from the
unbiased equilibrium in the extensive-form game,
completing the proof.

\section{Additional Experiments}

Beyond the experiments presented in the main text, we conduct a broader set of additional experiments to further validate our theoretical results and examine the influence of noise. These include extensive comparisons across multiple configurations of the regularized RD baselines with varying regularization strengths and update frequencies. The results consistently reveal an inherent difference in the contraction paths of the two backbone dynamics, which grants BNN greater adaptability under dynamic environments. Moreover, the experiments under varying noise levels clearly illustrate the emergence of the predicted noise floor.
\begin{table}[h!]
\centering
\caption{Summary of experimental figures.}
\begin{tabular}{c|c|c|c}
\hline
\textbf{Figure} & \textbf{Category} & \textbf{Game} & \textbf{Noise} \\
\hline
Fig.~1 & Stationary & BRPS / BRPS-W & True \\
Fig.~2 & Nonstationary (cont. \& direct) & BRPS & True \\
Fig.~3 & Nonstationary (cont. \& direct) & BRPS & Noise \\
Fig.~4 & Stationary & Leduc Poker & True \& Noise \\
Fig.~5 & Stationary & Kuhn Poker & True \& Noise \\
Fig.~6 & Nonstationary (cont. change) & Kuhn Poker & Noise \\
Fig.~7 & Nonstationary (direct change) & Kuhn Poker & Noise \\
\hline
\end{tabular}
\end{table}

\begin{figure}[b]
    \centering
    \includegraphics[width=\linewidth]{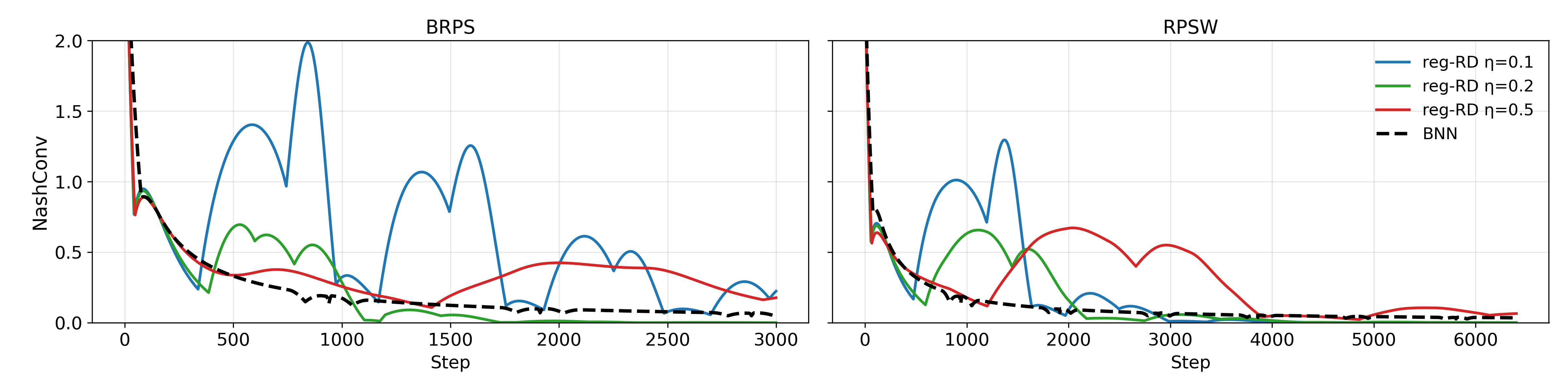}
    \caption{Stationary BRPS and BRPS-W without noise.}
\end{figure}
\begin{figure}[b]
    \centering
    \includegraphics[width=\linewidth]{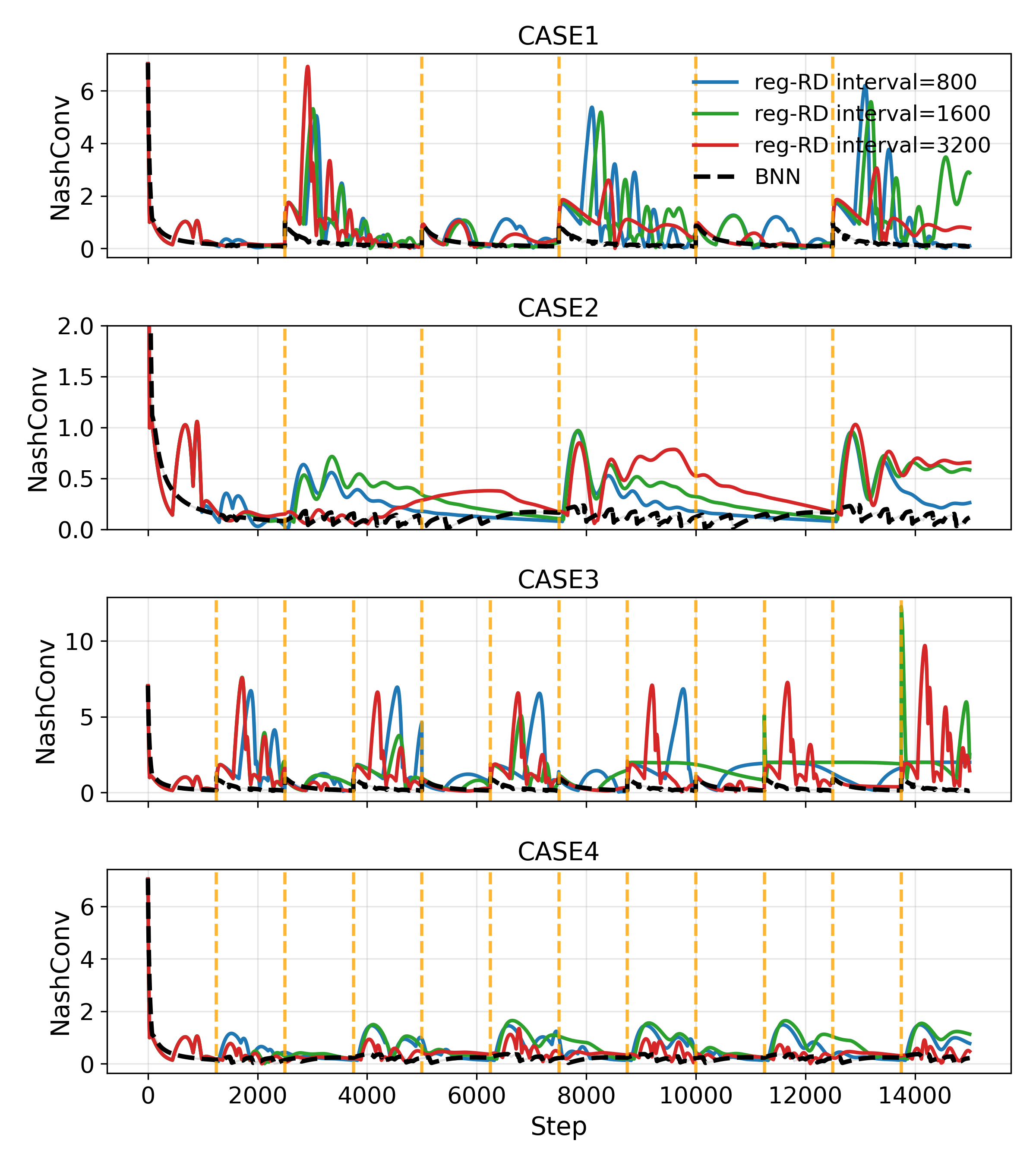}
    \caption{Nonstationary BRPS without noise.}
\end{figure}

\begin{figure}[b]
    \centering
    \includegraphics[width=\linewidth]{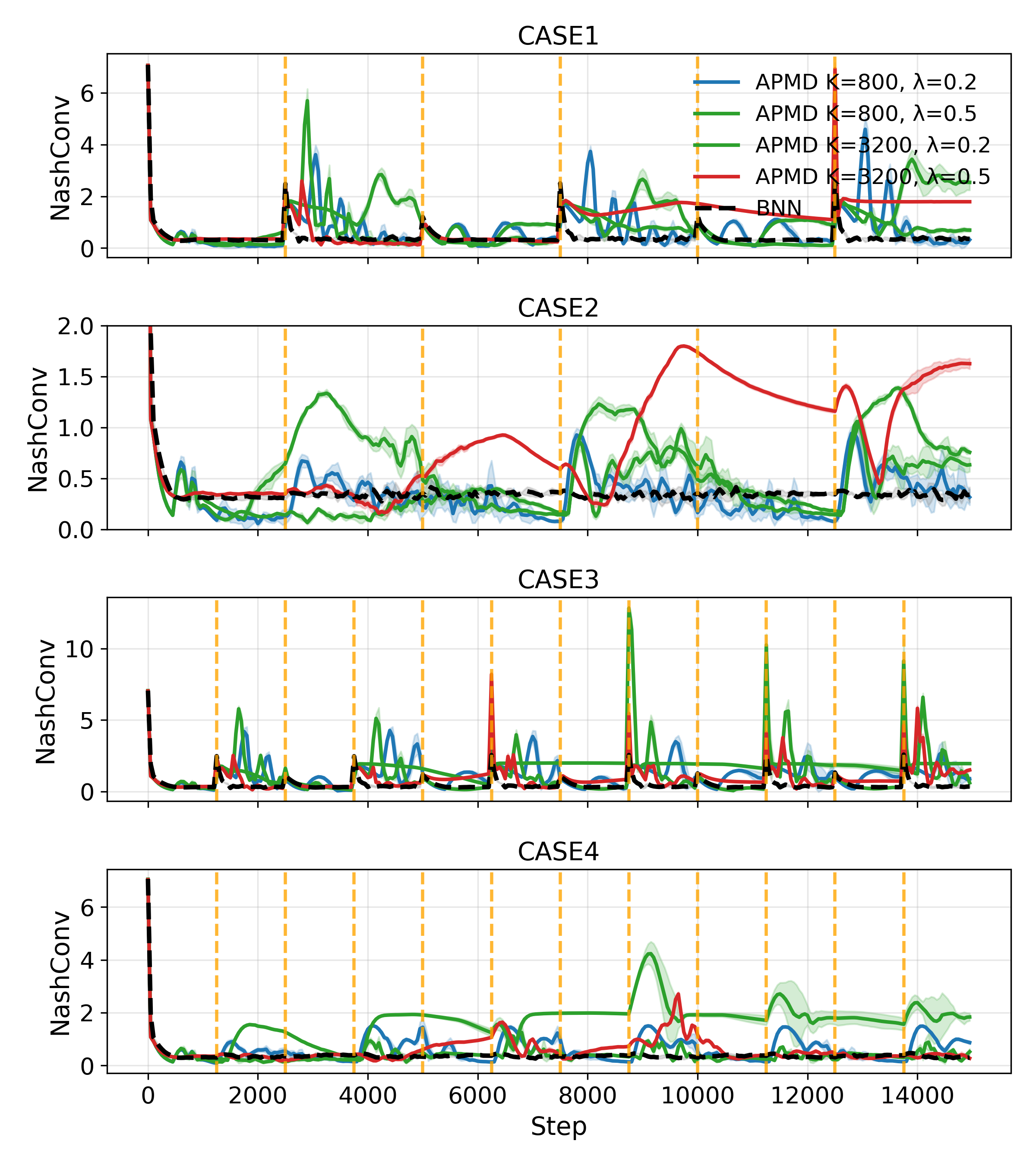}
    \caption{Nonstationary BRPS with different noise.}
\end{figure}

\begin{figure}[b]
    \centering
    \includegraphics[width=\linewidth]{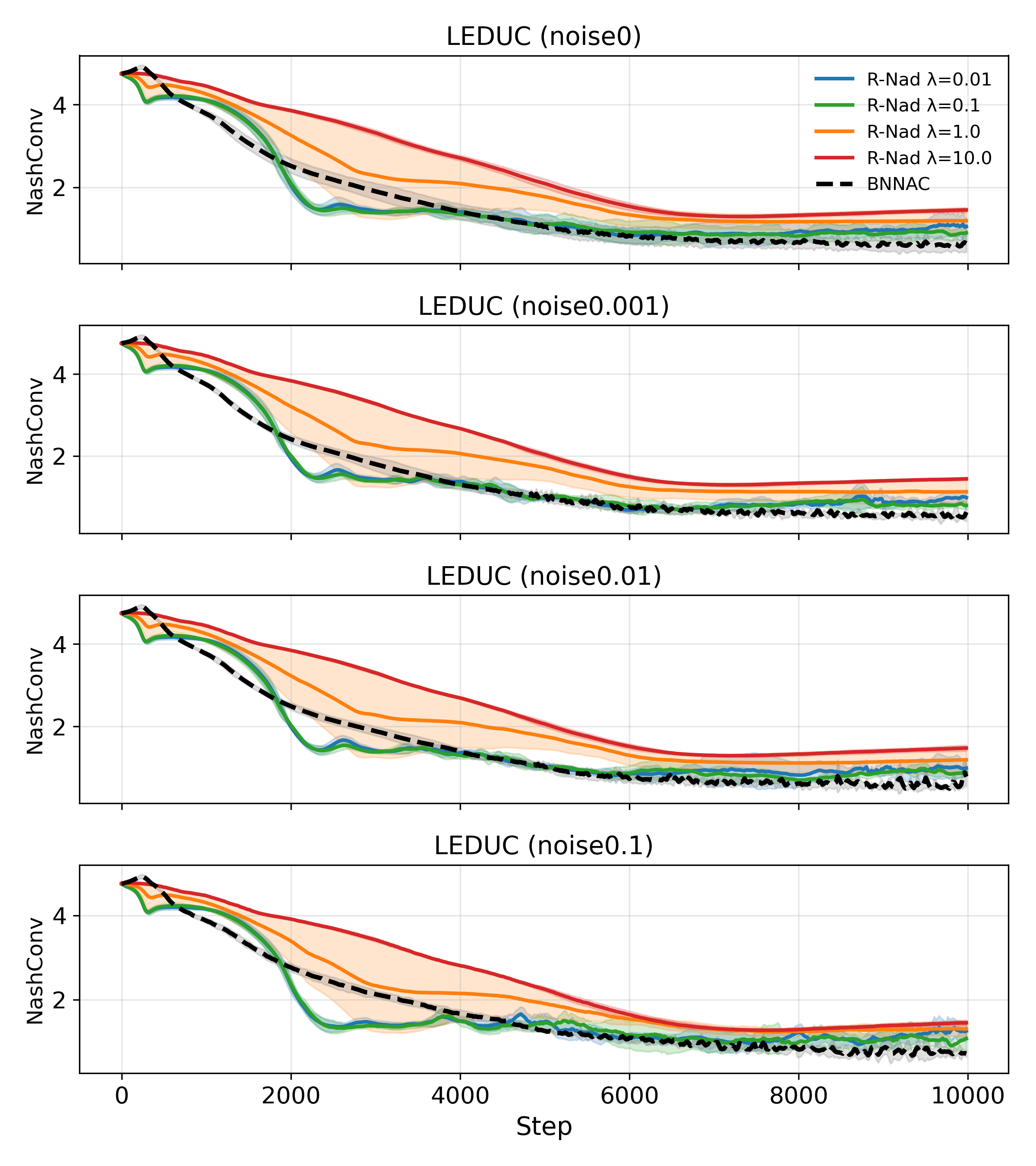}
    \caption{Stationary Leduc Poker with and without noise.}
\end{figure}
\begin{figure}[b]
    \centering
    \includegraphics[width=\linewidth]{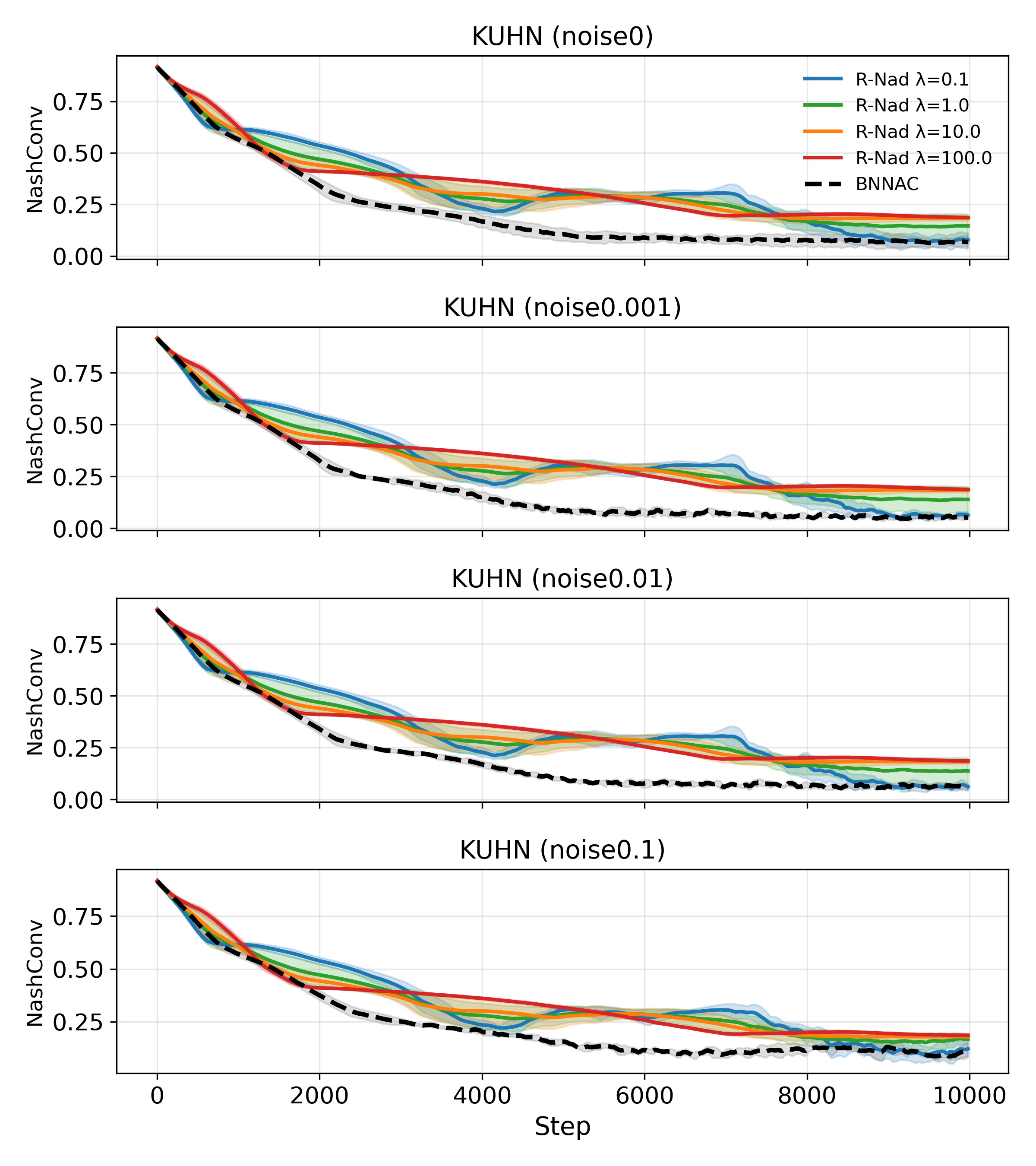}
    \caption{Stationary Kuhn Poker with and without noise.}
\end{figure}

\begin{figure}[b]
    \centering
    \includegraphics[width=\linewidth]{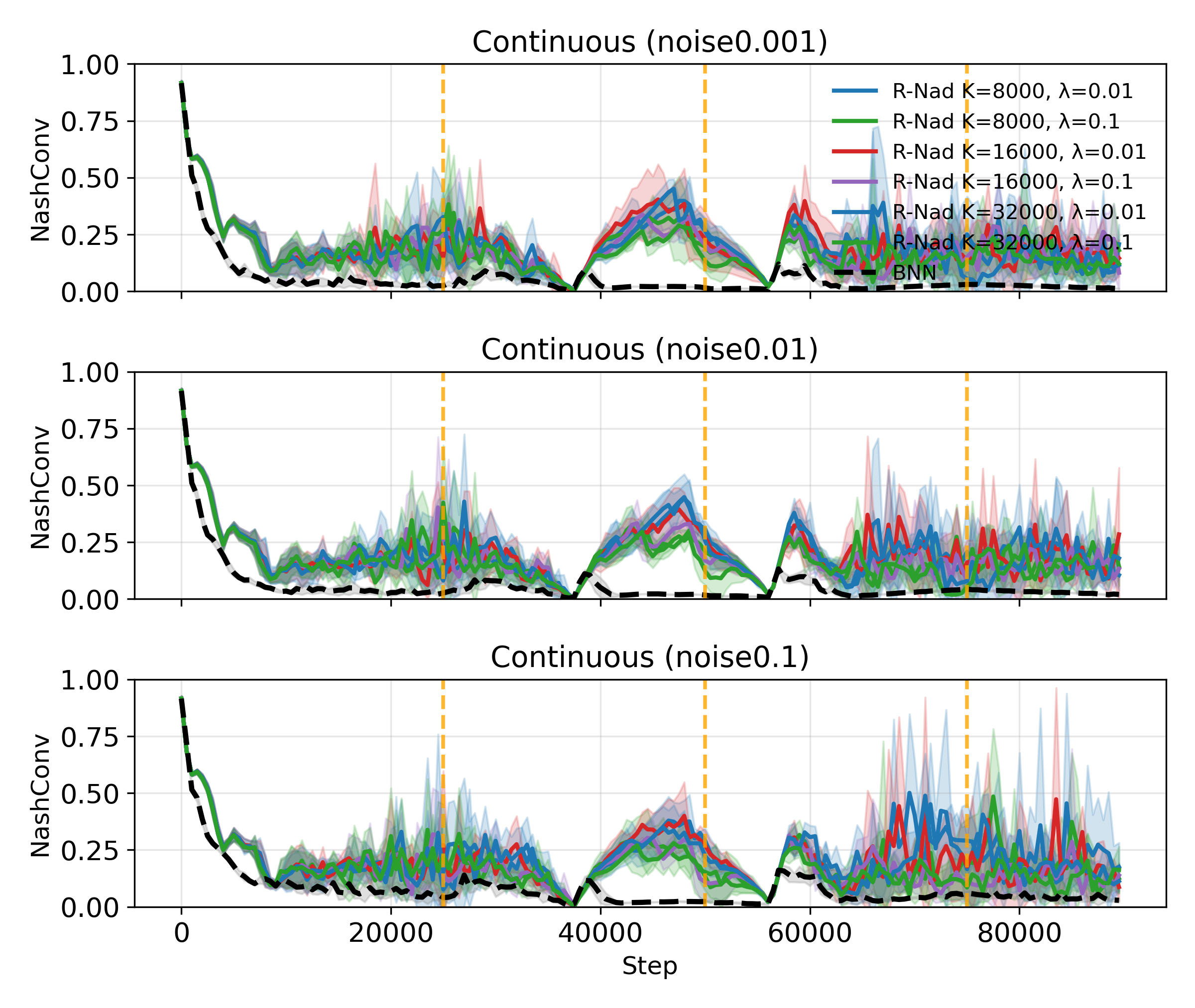}
    \caption{Nonstationary Kuhn Poker (continuous change) with different noise.}
\end{figure}
\begin{figure}[b]
    \centering
    \includegraphics[width=\linewidth]{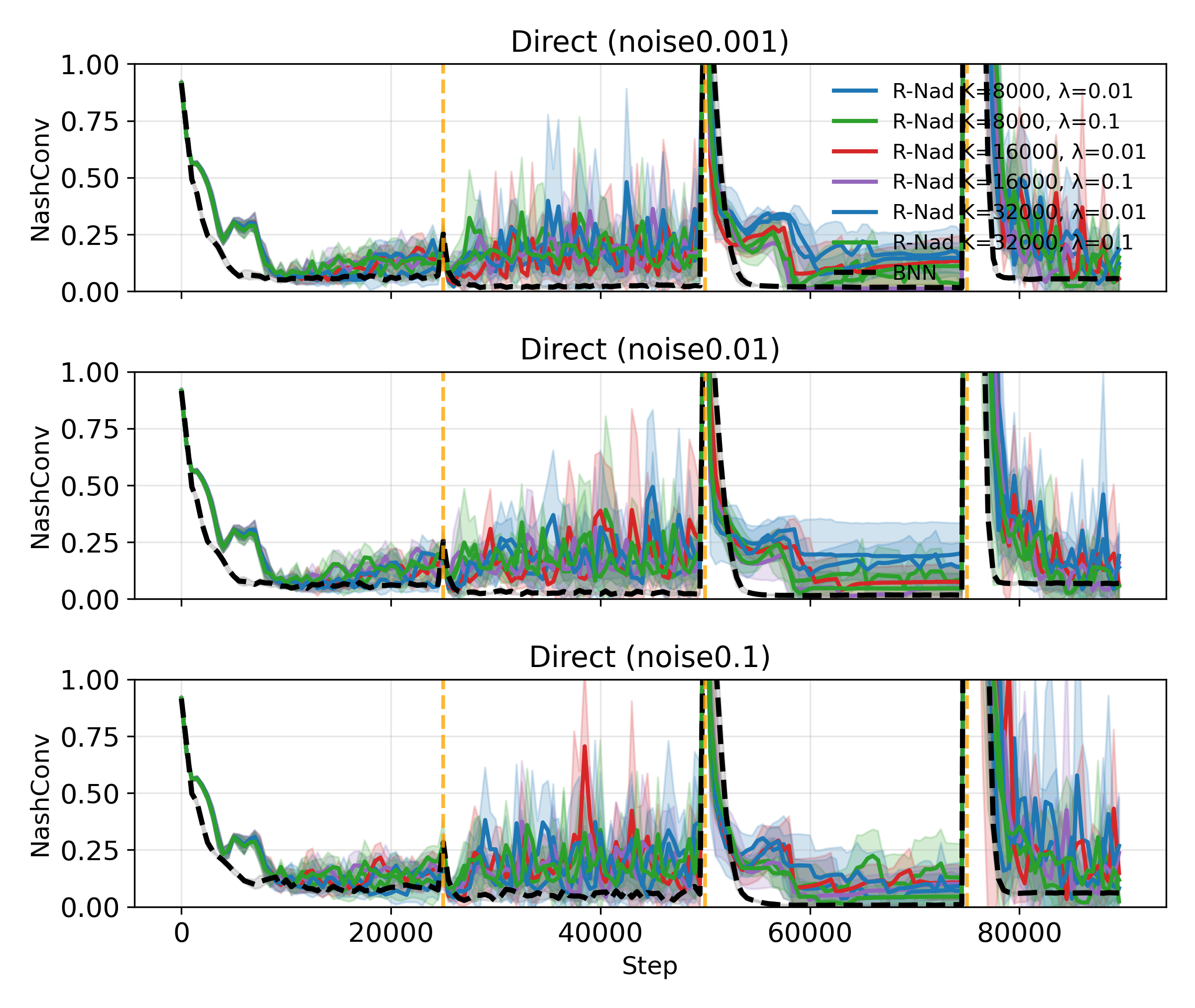}
    \caption{Nonstationary Kuhn Poker (direct change) with different noise.}
\end{figure}